\documentclass{aa}
\usepackage{psfig}
\usepackage{graphicx}

\newcommand{\gtrsim}{\mathrel{\hbox{\rlap{\lower.55ex \hbox {$\sim$}}
                   \kern-.3em \raise.4ex \hbox{$>$}}}}
\newcommand{\lesssim}{\mathrel{\hbox{\rlap{\lower.55ex \hbox {$\sim$}}
                   \kern-.3em \raise.4ex \hbox{$<$}}}}

\begin{document}

   \title{X-ray bursts at extreme mass accretion rates from GX\,17+2}

   \author{E.~Kuulkers
           \inst{1,2}
	   \and
	   J.~Homan
	   \inst{3}\fnmsep\thanks{Present address: Osservatorio Astronomico di Brera,
Via E.~Bianchi 46, I-23807 Merate (LC), Italy}
	   \and
           M.~van der Klis
           \inst{3}
	   \and
           W.H.G.~Lewin
           \inst{4}
           \and
	   M.~M\'endez
           \inst{1}
          }

   \offprints{Erik Kuulkers at SRON}

   \institute{
	      SRON National Institute for Space Research,
	      Sorbonnelaan 2, 3584 CA Utrecht, The Netherlands\\
              \email{E.Kuulkers@sron.nl, M.Mendez@sron.nl}
              \and
              Astronomical Institute, Utrecht University, 
              P.O.\ Box 80000, 3508 TA Utrecht, The Netherlands
              \and
	      Astronomical Institute ``Anton Pannekoek'', University of
	      Amsterdam, and Center for High Energy Astrophysics, Kruislaan 403,
              1098 SJ Amsterdam, The Netherlands\\
              \email{homan@merate.mi.astro.it, michiel@astro.uva.nl}
              \and
              Department of Physics and Center for Space Research,
              Massachusetts Institute of Technology, Cambridge, MA 02138, USA\\
              \email{lewin@space.mit.edu}
             }

   \date{Received --; accepted --}

   \titlerunning{X-ray bursts in GX\,17+2}

   \abstract{
We report on ten type~I X-ray bursts originating from GX\,17+2 in data obtained with the 
RXTE/PCA in 1996--2000. Three bursts were short in duration ($\sim$10\,s), whereas the others
lasted for $\sim$6--25\,min. All bursts showed spectral softening during their decay. There is no evidence
for high-frequency ($>$100\,Hz) oscillations at any phase of the bursts. 
We see no correlations
of the burst properties with respect to the persistent X-ray spectral properties, suggesting
that in GX\,17+2 the properties of the bursts do not correlate with inferred mass accretion rate. 
The presence of short bursts in GX\,17+2 (and similar bright X-ray sources) is not accounted for
in the current X-ray bursts theories at the high mass accretion rates encountered in this source.
~\\
We obtain satisfactory results if we model the burst emission with a black body, 
after subtraction of the persistent pre-burst emission. The two-component spectral model does not fit
the total burst emission whenever there is a black-body component present
in the persistent emission. We conclude that in those cases the black-body contribution from the 
persistent emission is also present during the burst. This implies that, contrary to previous suggestions,
the burst emission does {\it not} arise from the same site as the persistent black-body emission.
The black-body component of the persistent emission is consistent with being produced in
an expanded boundary layer, as indicated by recent theoretical work. 
~\\
Five of the long bursts showed evidence of radius expansion of the neutron star
photosphere (independent of the spectral analysis method used), 
presumably due to the burst luminosity reaching the Eddington value.
When the burst luminosity is close to the Eddington value, slight deviations from pure black-body radiation
are seen at energies below $\simeq$10\,keV. Similar deviations have been seen during (long) X-ray bursts from
other sources; they can not be explained by spectral hardening models.
~\\
The total persistent flux just before and after the radius expansion bursts 
is inferred to be up to a factor of 2 higher than the net peak flux of the burst. 
If both the burst and persistent emission are radiated isotropically,
this would imply that the persistent emission is up to a factor of 2 higher than the Eddington 
luminosity. This is unlikely and we suggest that the persistent luminosity is close to the 
Eddington luminosity and that the burst emission is (highly) anisotropic ($\xi$$\sim$2).
Assuming that the net burst peak fluxes equal the Eddington limit, 
applying standard burst parameters (1.4\,M$_{\sun}$ neutron star,
cosmic composition, electron scattering opacity appropriate for high temperatures), and
taking into account gravitational redshift and spectral hardening, 
we derive a distance to GX\,17+2 of $\sim$8\,kpc, with an
uncertainty of up to $\sim$30\%.
  }

   \maketitle

\section{Introduction}

X-ray bursts were discovered in 1975 from the 
source 4U\,1820$-$30 (Grindlay \&\ Heise 1975; Grindlay et al.\ 1976).
It was realized soon thereafter that these were thermo-nuclear runaway events on the 
surface of neutron stars (Woosley \&\ Taam 1976; Maraschi \&\ Cavaliere 1977).
Another kind of X-ray bursts was found (together with the above type of bursts) from 
MXB\,1730$-$355 (later referred to as the Rapid Burster), which were suggested to be due to 
accretion instabilities. The former and latter kind of bursts were then dubbed type~I
and type~II, respectively (Hoffman et al.\ 1978a).

The main characteristics of type~I bursts (for a review see Lewin et al.\ 1993) are:
sudden and short ($\simeq$1\,s) increase in the X-ray flux, exponential decay light curve,
duration of the order of seconds to minutes, softening during the decay (attributed to
cooling of the neutron star surface), (net) burst spectra reasonably well described by black-body 
emission from a compact object with $\simeq$10\,km radius and temperature of $\simeq$1--2\,keV,
and total energies ranging from $\simeq$10$^{39}$ to 10$^{40}$\,erg.
When the luminosity during the burst reaches the Eddington 
limit (i.e., when the pressure force due to radiation balances the gravitational force), 
the neutron star photosphere expands.
Since L$_{\rm b}\propto {\rm R}^2{\rm T_{\rm eff}}^4$, when the radius of the photosphere, R, expands, the
effective temperature, T$_{\rm eff}$, drops, with the burst luminosity, L$_{\rm b}$, 
being constant (modulo gravitational
redshift effects with changing R) at the Eddington limit, L$_{\rm Edd}$. 
Bursts during their radius expansion/contraction phase are therefore recognizable by 
an increase in the inferred radius with a simultaneous decrease in the observed temperature, while
the observed flux stays relatively constant.

The emission from a (hot) neutron star is not expected to be perfectly Planckian, however
(van Paradijs 1982; London et al.\ 1984, 1986; see also Titarchuk 1994; Madej 1997, and 
references therein). This is mainly due to the effects of electron scattering in the 
neutron star atmosphere, deforming the original X-ray spectrum.
This results in a systematic difference between the effective 
temperature (as would be measured on Earth), T$_{\rm eff,\infty}$, 
and the temperature as obtained from the spectral fits, T$_{\rm bb}$
(also referred to as `colour' temperature, see e.g.\ Lewin et al.\ 1993).
In general, the deviations from a Planckian distribution will
depend on several parameters, such as temperature, elemental abundance, neutron star mass and radius. 
The hardening factor, T$_{\rm bb}$/T$_{\rm eff,\infty}$, has been determined through numerical calculations
by various people and its value is typically around 1.7.
When the burst luminosity approaches the Eddington limit the deviations from a black-body become larger,
and so does the spectral hardening (T$_{\rm bb}$/T$_{\rm eff,\infty}$$\sim$2, Babul \&\ Paczy\'nski 1987;
Titarchuk 1988). During extreme radius expansion phases, 
however, this trend may break down and T$_{\rm bb}$/T$_{\rm eff,\infty}$$<$1 (Titarchuk 1994).
Attempts have been made to determine the spectral hardening from the observed cooling tracks,
but conclusions are still rather uncertain (e.g.\ Penninx et al.\ 1989; Smale 2001).
As a result, the interpretation
of X-ray bursts spectra has remained uncertain and constraints on the 
mass-radius relationship for neutron stars elusive.

Type~I X-ray burst theory predicts three different regimes in mass accretion rate (\.M) for
unstable burning 
(Fujimoto et al.\ 1981, Fushiki \&\ Lamb 1987; see also Bildsten 1998, 2000, 
Schatz et al.\ 1999, and references therein; note that values of critical \.M depend on
metallicity, and on assumed core temperature and mass of the neutron star):
\begin{itemize}
\item[1)] low accretion rates; $10^{-14}$\,${\rm M}_{\sun}$\,${\rm yr}^{-1}\lesssim \dot{\rm M} \lesssim 2\times 10^{-10}$\,${\rm M}_{\sun}$\,${\rm yr}^{-1}$: 
mixed H/He burning triggered by thermally unstable H ignition
\item[2)] intermediate accretion rates; $2\times 10^{-10}$\,${\rm M}_{\sun}$\,${\rm yr}^{-1} \lesssim \dot{\rm M} \lesssim 4$--$11\times 10^{-10}$\,${\rm M}_{\sun}$\,${\rm yr}^{-1}$: 
pure He shell ignition after steady H burning
\item[3)] high accretion rates; 4--$11\times 10^{-10}$\,${\rm M}_{\sun}$\,${\rm yr}^{-1} \lesssim \dot{\rm M} \lesssim 2\times10^{-8}$\,${\rm M}_{\sun}$\,${\rm yr}^{-1}$: 
mixed H/He burning triggered by thermally unstable He ignition
\end{itemize}
H and He are burning stably in a mixed H/He environment for very low and very high values of 
$\dot{\rm M}$, i.e.\ $\dot{\rm M}$ below $\sim$$10^{-14}$\,${\rm M}_{\sun}$\,${\rm yr}^{-1}$ and above
$\sim$$2\times10^{-8}$\,${\rm M}_{\sun}$\,${\rm yr}^{-1}$ (close to the critical Eddington \.M).
During pure helium flashes the fuel is burned rapidly, and such bursts therefore last only 5--10\,s.
This gives rise to a large energy release in a short time, which causes the bursts often 
to reach the Eddington limit, leading to photospheric radius expansion.  
Bursts with unstable mixed H/He burning release their energies
on a longer, 10--100\,s, timescale, due to the long series of $\beta$ decays in the rp-process
(see e.g.\ Bildsten 1998, 2000).

The Z sources (Hasinger \&\ van der Klis 1989) are a group of sources inferred to persistently accrete 
near the Eddington limit. In a colour-colour diagram they trace out a Z-like shape,
with the three limbs of the Z (historically) referred to as the horizontal branch (HB), normal
branch (NB) and flaring branch (FB), from top to bottom. 
\.M is inferred to increase from sub-Eddington at the HB, near-Eddington at 
the NB to super-Eddington at the FB (e.g.\ Hasinger 1987; Lamb 1989; 
Hasinger et al.\ 1990). According to the burning regimes
outlined above these sources should exhibit long ($\gtrsim$10--100\,s) type I X-ray bursts,
at least on the HB and NB. However, 
of the Z sources, only GX\,17+2 and Cyg\,X-2 show (infrequent) bursts
(Kahn \&\ Grindlay 1984; Tawara et al.\ 1984c; Sztajno et al.\ 1986; Kuulkers et al.\ 1995, 1997; 
Wijnands et al.\ 1997; Smale 1998), indicating that most of the material is burning stably.
This is in contrast to the above theoretical expectations for high \.M.
Moreover, the bursts in Cyg\,X-2 are short ($\simeq$5\,s), whereas GX\,17+2 shows both short
($\simeq$10\,s) and long ($\gtrsim$100\,s) bursts.
One of the bursts of Cyg\,X-2 showed a radius expansion phase (Smale 1998); this burst
clearly bears all the characteristics of a He flash (regime 2), whereas the neutron star is inferred to 
accrete at near-Eddington rates. The short duration bursts in GX\,17+2 also hint to a
He flash origin, whereas the long duration bursts hint to unstable mixed H/He burning
(regime 3; see also van Paradijs et al.\ 1988). 
A study of X-ray bursts from Cyg\,X-2 and GX\,17+2 observed by EXOSAT showed 
no correlation of the burst properties 
with position in the Z, although the number of bursts observed from GX\,17+2 was small 
(Kuulkers et al.\ 1995, 1997).

The {\em Proportional Counter Array} (PCA) onboard the 
{\em Rossi X-ray Timing Explorer} (RXTE)
combines high throughput using a large collecting area (maximum of $\simeq$6500\,cm$^2$)
with the ability to label photons down to a time resolution of $\mu$s. This is ideal to
study short events like X-ray bursts, especially during the start of
the burst. Such studies may provide more insight in the
properties of X-ray bursts which occur at these extreme mass accretion rates.
Analysis of X-ray bursts in persistent sources at (relatively) high inferred mass accretion rates
(typically $\gtrsim$0.2\,\.M$_{\rm Edd}$) observed
with the RXTE/PCA were presented for one burst seen with Cyg\,X-2 (Smale 1998) and one seen in GX\,3+1
(Kuulkers \&\ van der Klis 2000). 
In this paper we present the first account of ten X-ray bursts from GX\,17+2 
observed by the RXTE/PCA during the period 1996--2000. For a description of the 
correlated X-ray timing and spectral
properties of GX\,17+2 using the same data set we refer to Homan et al.\ (2001).

\section{Observations and Analysis}

\begin{table*}
\caption{RXTE observation log of GX\,17+2$^a$}
\begin{tabular}{ccccccccc}
\hline
\multicolumn{1}{c}{Year} & \multicolumn{3}{c}{Start (UT)} & \multicolumn{3}{c}{End (UT)} &
\multicolumn{1}{c}{t$_{\rm exp}$$^b$} & \multicolumn{1}{c}{$\#$ of} \\
\multicolumn{1}{c}{~} & \multicolumn{3}{c}{~} & \multicolumn{3}{c}{~} &
\multicolumn{1}{c}{(ksec)} & \multicolumn{1}{c}{bursts} \\
\hline
1996 & feb & 07 & 13:27 & feb & 09 & 00:02 & 58 & 1 \\
1997 & feb & 02 & 19:13 & feb & 27 & 03:34 & 59 & 1 \\
1997 & apr & 01 & 19:13 & apr & 04 & 23:26 & 35 & 0 \\
1997 & jul & 27 & 02:13 & jul & 28 & 00:33 & 43 & 0 \\
1998 & aug & 07 & 06:40 & aug & 08 & 23:40 & 71 & 1 \\
1998 & nov & 18 & 06:42 & nov & 20 & 13:31 & 86 & 2 \\
1999 & oct & 03 & 02:43 & oct & 12 & 07:05 & 298 & 5 \\
2000 & mar & 31 & 12:15 & mar & 31 & 16:31 & 7 & 0 \\
\hline
\multicolumn{9}{l}{$^a$\,Between Feb 2--27, 1997 observations every 3--6~days.} \\
\multicolumn{9}{l}{$^b$\,Total effective exposure time.} \\
\end{tabular}
\end{table*}

The PCA (2--60\,keV; Bradt et al.\ 1993) onboard RXTE observed GX\,17+2 
various times during the mission. Up to now a total of 657\,ksec of useful data
has been obtained. A log of these observations is given in Table~1.
During the observations in 1996--1998 all five proportional counter units (PCUs) were active,
whereas in 1999 and 2000 only three units were active.
The high voltage settings of the PCUs have been altered three times during the mission
(so-called gain changes), which modified the response of the detectors. 
These changes therefore mark four periods, called gain epochs.
The observations were done in two standard modes:
one with relatively high spectral resolution (129 energy channels covering the 
whole PCA energy band) every 16\,sec, the {\sc Standard~2} mode, the other
having no spectral information providing the intensity in the whole PCA energy band
at a moderate time resolution of 0.125\,s, the {\sc Standard~1} mode. 
Additionally, data were recorded in various high time resolution ($\leq$2\,ms) modes 
that ran in parallel to the {\sc Standard} modes, and that recorded photons  
within a specific energy band with either low spectral resolution (B-modes or E-modes) or
no spectral resolution (SB-modes). The B-, E- and
SB-modes used here combined the information from all layers of all active PCUs together.
During the 1996 observations a B- and E-mode were available, giving 16 and 64 energy bands, covering channels
0--49 and 50--249, at 2\,ms and 125\,$\mu$s,
respectively. For most of the observations in 1997 four SB modes 
covering the total PCA energy range were available. 
In the 1998, 1999, and 2000 observations two SB-modes covering channels 0--13 and
14--17 at 125$\mu$s, and an E-mode giving 64 energy bands covering channels 18--249 at
16\,$\mu$s time resolution were available.

For the spectral analysis of the persistent emission we used the {\sc Standard~2} data.
We accumulated data stretches of 96\,s just before the burst, 
combining the PCUs which were operating at that time. 
In order to study the spectral properties of the bursts we used two approaches.
Spectra during the bursts were determined every 0.25\,sec for the first $\simeq$20\,s 
of the burst. For the short bursts this means the whole duration of the burst.
For the long bursts we also used the {\sc Standard~2} data to create spectra at 16\,s intervals,
for evaluating the remainder of these bursts.
Since no high time resolution spectral data were available during the 1997 observations
(only 4 SB-modes), 
only the {\sc Standard~2} data were used to study the spectral properties of the 
burst from this observation. 
All spectra were corrected for background and dead-time using the procedures 
supplied by the RXTE Guest Observer 
Facility\footnote{http://heasarc.gsfc.nasa.gov/docs/xte/recipes/cook\_book.html.}.
A systematic uncertainty of 1\%\ in the count rate spectra was taken into account.
For our spectral fits we confined ourselves to the energy range of 
3--20\,keV, which is best calibrated.
The hydrogen column density, N$_{\rm H}$, towards GX\,17+2 was fixed
to that found by the {\em Einstein} SSS and MPC measurements (2$\times$10$^{22}$\,atoms\,cm$^{-2}$,
Christian \&\ Swank 1997; see also Di Salvo et al.\ 2000). In all cases we included
a Gaussian line (see Di Salvo et al.\ 2000) fixed at 6.7\,keV, with a fixed line width
of 0.1\,keV. One sigma confidence errors were determined using $\Delta\chi^2=1$.

\begin{figure}
 \resizebox{\hsize}{!}{\includegraphics[angle=-90, clip, bb=78 38 576 550]{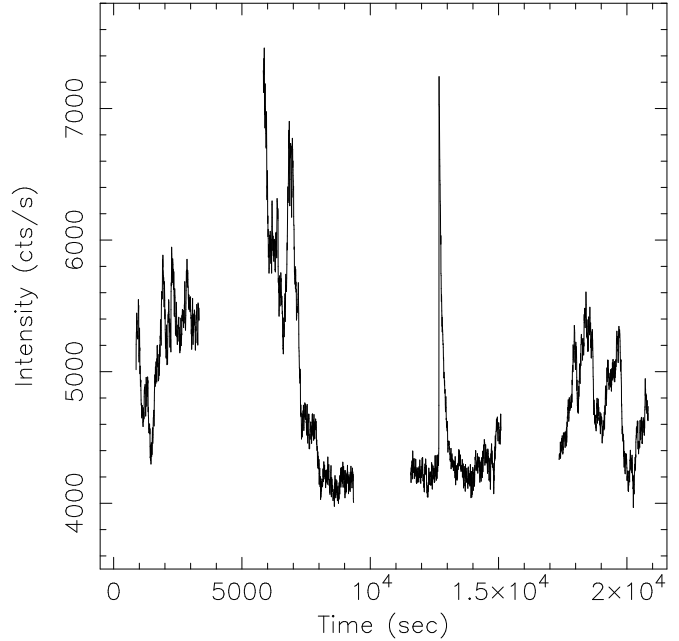}}
\caption{{\sc Standard~1} light curve of the GX\,17+2 observations during October 10, 1999,
at a time resolution of 5\,s. Time zero corresponds to 05:39:27 (UTC). 
No corrections for background and dead-time have been applied.
Clearly, the source varies on the same time scale (and faster) as the burst which started at
${\rm T}=12680$\,s. The source was in the FB and the lower part of the NB during the 
observations.}
\label{lc_b10}
\end{figure}

Large amplitude, high coherence brightness oscillations have been observed
during various type~I X-ray bursts in other low-mass X-ray binaries (LMXBs; see e.g.\ Strohmayer 1998, 2001).
We searched all the bursts from GX\,17+2 for such
oscillations. Using the high time resolution modes we performed Fast
Fourier Transforms to produce power spectra with a Nyquist
frequency of 2048 Hz. This was done in the total energy band (2--60\,keV) 
and in a high energy band ranging from $\simeq$8\,keV to 
$\simeq$20\,keV. For the 1996 observations, however, the high time resolution mode only covered
the 13.5--60\,keV range; the 2--60\,keV power spectra had a Nyquist frequency of
256\,Hz. In all cases, our searches were carried out on power spectra of 0.25\,s or 2\,s in
duration. To increase the sensitivity for cases 
where burst oscillations are only present for a period of time comparable or shorter than the length of the
power spectrum, we `oversampled' 
the data by factors of 2 and 8, respectively, by 
taking the start time of the next data segment to be
0.125\,s and 0.25\,s later than that of the previous one, instead of
0.25\,s and 2\,s (i.e.\ we use overlapping data segments) . 

\begin{table*}
\caption{Bursts and burst-like events (flares) in GX\,17+2}
\begin{tabular}{ccccccccccccc}
\hline
\multicolumn{1}{c}{event$^a$} & \multicolumn{4}{c}{Start Time (UT)} & 
\multicolumn{1}{c}{E$^b$} &
\multicolumn{1}{c}{PCUs$^c$} &  \multicolumn{1}{c}{t$_{\rm dur}$$^d$} &
\multicolumn{1}{c}{t$_{\rm rise}$$^e$} & \multicolumn{1}{c}{t$_{\rm fr}$$^f$} &
\multicolumn{1}{c}{t$_{\rm exp}$$^g$} & \multicolumn{1}{c}{$\chi^2_{\rm red}$/dof$^h$} & 
\multicolumn{1}{c}{branch$^i$} \\
\hline
 b1 & 1996 & feb & 08 & 16:17:12 & 1 & 5          & 10     & 1.22 & 0.35 $\pm$ 0.05 & 1.83 $\pm$ 0.08  & 1.1/130 & mNB \\
 b2 & 1997 & feb & 08 & 02:36:34 & 3 & 5          & $>$360 & 1.19 & 0.34 $\pm$ 0.08 & 248 $^{+4}_{-9}$ & 2.5/49  & mNB \\
 b3 & 1998 & aug & 07 & 13:15:50 & 3 & 5          & 10     & 0.53 & 0.27 $\pm$ 0.09 & 2.55 $\pm$ 0.24  & 1.0/131 & lNB \\
 b4 & 1998 & nov & 18 & 08:51:26 & 3 & 5          & 1000   & 1.34 & 0.61 $\pm$ 0.04 & 197 $\pm$ 2      & 3.7/147 & SV \\
 b5 & 1998 & nov & 18 & 14:37:30 & 3 & 5          & 10     & 0.41 & 0.54 $\pm$ 0.04 & 2.06 $\pm$0.13   & 1.1/85  & mNB \\
 b6 & 1999 & oct & 03 & 15:36:32 & 4 & 3 (0,2,3)  & 1600   & 0.41 & 0.19 $\pm$ 0.04 & 274 $\pm$ 3      & 2.0/242 & lHB \\
 b7 & 1999 & oct & 05 & 23:41:43 & 4 & 3 (0,2,4)  & 500    & 0.56 & 0.30 $\pm$ 0.03 & 77.3 $\pm$ 1.2   & 1.9/104 & uNB \\
 b8 & 1999 & oct & 06 & 11:10:33 & 4 & 3 (0,2,3)  & 500    & 1.66 & 0.13 $\pm$ 0.02 & 70.2 $\pm$ 1.4   & 1.2/57  & lHB \\
 b9 & 1999 & oct & 09 & 12:34:24 & 4 & 3 (0,2,3)  & 500    & 0.13 & 0.16 $\pm$ 0.02 & 76.4 $\pm$ 1.5   & 3.1/66  & uNB \\
b10 & 1999 & oct & 10 & 09:10:47 & 4 & 3 (0,2,3)  & 700    & 0.72 & 0.20 $\pm$ 0.04 & 115 $\pm$ 3      & 3.7/57  & lNB \\
\multicolumn{13}{l}{~} \\
 f1 & 1996 & feb & 07 & 03:39:11 & 1 & 5          & 10     & 1.03 & 1.12 $\pm$ 0.07 & 2.98 $\pm$ 0.49  & 1.3/88  & mFB \\
 f2 & 1998 & nov & 19 & 14:38:24 & 3 & 5          & 10     & 0.41 & 0.33 $\pm$ 0.04 & 1.72 $\pm$ 0.35  & 1.1/28  & uFB \\
 f3 & 1998 & nov & 20 & 00:44:43 & 3 & 5          & 10     & 0.75 & 0.45 $\pm$ 0.06 & 1.85 $\pm$ 0.25  & 1.1/55  & uFB \\
 f4 & 1999 & oct & 11 & 08:55:30 & 4 & 3 (0,2,3)  & 10     & 5.0  & 1.06 $\pm$ 0.20 & 2.18 $\pm$ 0.22  & 1.1/148 & lFB \\
\hline
\multicolumn{13}{l}{$^a$\,Designation used in text.} \\
\multicolumn{13}{l}{$^b$\,RXTE gain epoch of the observation.} \\
\multicolumn{13}{l}{$^c$\,Numbers of active PCUs; if $<$5 the PCU unit numbers are given between brackets.} \\
\multicolumn{13}{l}{$^d$\,Approximate duration in sec.} \\
\multicolumn{13}{l}{$^e$\,Time (sec) for the count rate to rise from 25\%\ to 90\%\ of the net-peak rate.} \\
\multicolumn{13}{l}{$^f$\,Duration of fast rise phase in sec, see text.} \\
\multicolumn{13}{l}{$^g$\,Exponential decay time in sec.} \\
\multicolumn{13}{l}{$^h$\,Goodness of the exponential fit to the decay part of the light curve.} \\
\multicolumn{13}{l}{$^i$\,Source position in the Z just before the event (see also Fig.~\ref{cd}); u, m, and l stand for upper, middle and lower,}\\
\multicolumn{13}{l}{$^{~}$\, respectively. SV means `soft' vertex, i.e.\ the NB/FB transition region.} \\
\end{tabular}
\end{table*}

\section{Results}

\subsection{Temporal behaviour}

\begin{figure*}
 \resizebox{12cm}{!}{\includegraphics[angle=-90, clip, bb=28 41 574 762]{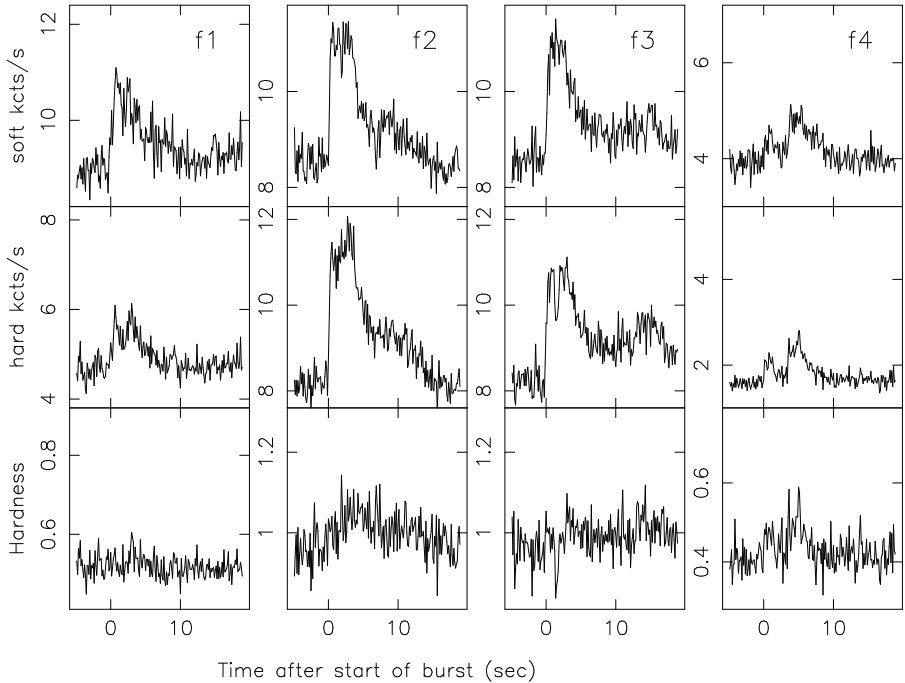}}
   \caption{The X-ray flare light curves at low (upper panels) and high (middle panels) 
energies and the corresponding hardness curves (lower panels), all at a time resolution 
of 0.125\,sec, for the four flares (f1--f4; see text). Time zero is the flare start time as given in 
Table~2. The low and high energy ranges are 
1.5--7.2\,keV and 7.2--19.7\,keV, respectively, for f1, 
1.9--6.2\,keV and 6.2--19.6\,keV, respectively, for f2 and f3,
and 2.1--7.1\,keV and 7.1--19.9\,keV, respectively, for 4. Hardness is defined as the 
count rate ratio of the high to low energy band.
No corrections for background and dead-time have been applied.}
   \label{plot_short_f1-f4}
\end{figure*}

\begin{figure*}
 \resizebox{12cm}{!}{\includegraphics[angle=-90, clip, bb=28 41 574 769]{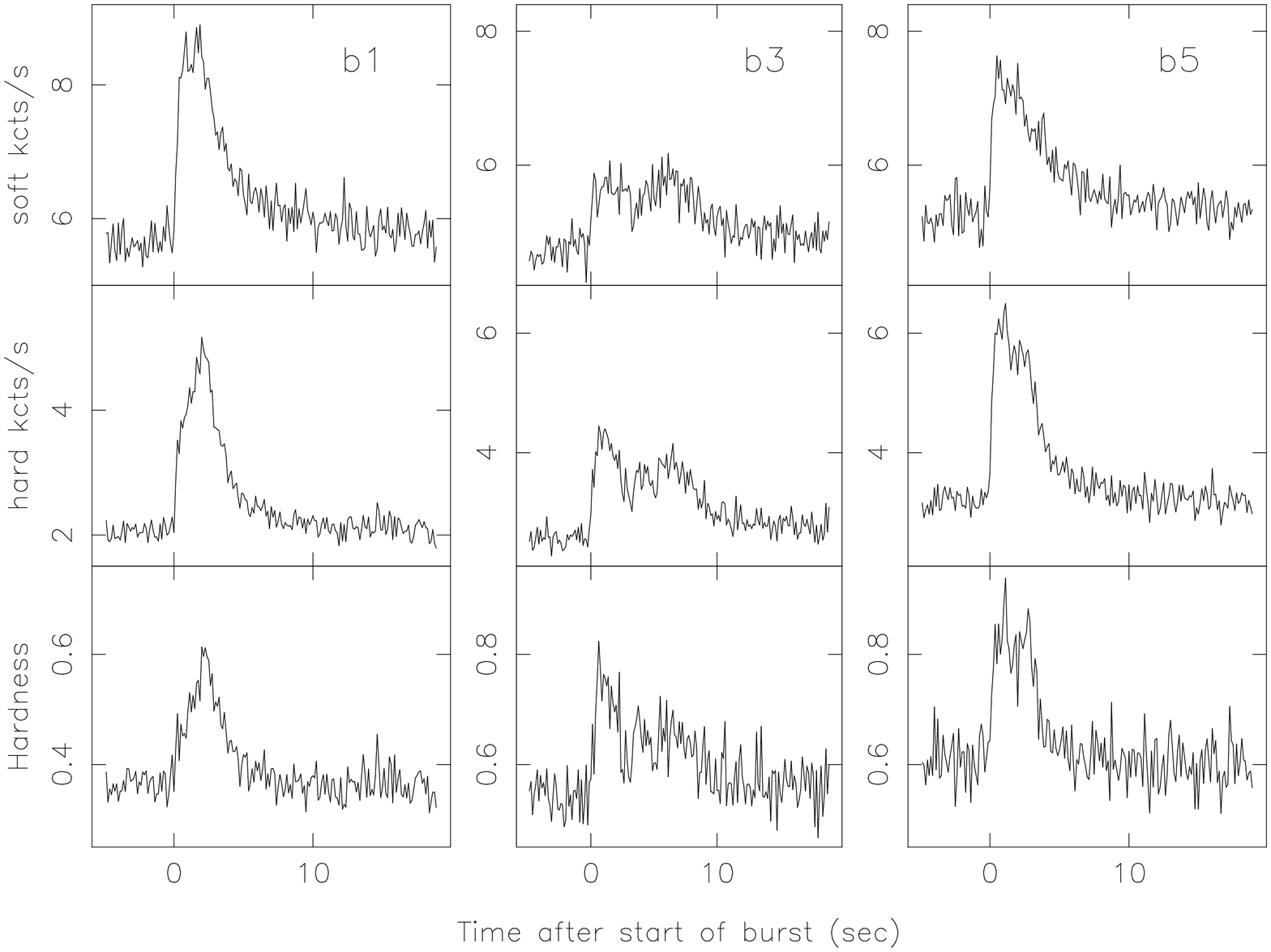}}
   \caption{Same as Fig.~\ref{plot_short_f1-f4}, but for the three short X-ray bursts
b1, b3, and b5. The low and high energy ranges are 
1--7.2\,keV and 7.2--19.7\,keV, respectively, for b1, whereas they are 
1.9--6.2\,keV and 6.2--19.6\,keV, respectively, for b3 and b5.}
   \label{plot_b1_b3_b5}
\end{figure*}

Inspection of all the {\sc Standard 1} light curves of GX\,17+2 at 1\,s time resolution
yielded a total of fourteen burst-like events (designated b1, b2, ..., b10, f1, ..., f4; the reason 
for different designations is the nature of the events, see below).
Seven events were rather short, $\simeq$10\,s (b1, b3, b5, f1--f4), whereas the other events
were long, $\simeq$360--1600\,s (b2, b4, b6--b10).
Table~2 gives the start times of these events, together with the gain epoch in which the
observations were done and the number of active PCUs. In that table we also give
the duration of the events as estimated by eye using the light curves in the total PCA energy band
and/or the hardness (ratio of count rates in two energy bands) 
curves, the rise time of the event as determined in two ways (see below),
and the exponential decay time as determined from 
a fit to the decay part of the {\sc Standard~1} light curves,
together with the goodness of this fit expressed in terms of 
reduced $\chi^2$, $\chi^2_{\rm red}$.
We note that the estimates of the durations are rather arbitrary, but 
robust methods such as the so-called T90 measurements\footnote{T90 is 
defined as the time it takes to observe 90\%\ of the total
background-subtracted counts in an event, starting and ending when 5\%\ and 95\%\ of 
the total background-subtracted counts have been observed.}
as used for $\gamma$-ray bursts (see Koshut et al.\ 1996, and references therein)
can not be applied, especially to the longer events, since the persistent emission
varies on the same time scale (or even fatser) as the event itself, see e.g.\ Fig.~\ref{lc_b10}.
The rise time of the events was determined as follows. We constructed light curves in the 
full PCA energy band with a time resolution of 1/32\,sec. We defined
t$_{\rm rise}$ as the time for the event to increase from 25\%\ to 90\%\ of the
net peak event rate (see e.g.\ Muno et al.\ 2000; van Straaten et al.\ 2001).
A detailed look at the light curves revealed that the onset of events b1--b4, b8, b10, and f2
consisted of a rapid rise phase and a subsequent slower rise to maximum.
In the events b5--b7, b9, f1, and f3 the count rate rapidly increased to maximum, 
with no subsequent slower rise. 
We therefore also fitted the pre-event phase, fast rise phase (and slow rise phase) with
a constant level and one (or two) linear functions, respectively, using the 
{\sc Standard~1} light curves.
We find that the total duration of the fast rise phase, t$_{\rm fr}$, 
is between 0.1 and 0.6\,sec for events b1--b10 and 0.4--1\,s for f1--f4 (Table 2). 
In the fits to the decay portion of the light curves for the 
short events we used the {\sc Standard~1} light curves at 0.125\,sec time resolution, while for the 
longer events we rebinned these light curves to a time resolution of 5\,sec.
For some of the events an exponential does not describe the decay very well;
this is probably due to the short term variations in the persistent emission.
For these events we fitted only the initial decay (first few seconds for f2 and f3, and first few 100\,s
for burst b9 and b10).

In Fig.~\ref{plot_short_f1-f4} we show the light curves of the four events f1--f4,
at low ($\lesssim$7\,keV) and high ($\gtrsim$7\,keV) energies, 
with the corresponding hardness (ratio of the count rates in the high and low energy band) curves,
all at a time resolution of 0.125\,s.
Although they have a fast rise and a longer decay (see also Table 2), they show small or no
variations in hardness. Time-resolved spectral analysis (like done in Sect.~3.2.3) 
confirmed this; no clear cooling during the decay can be discerned. We can therefore
not classify these events as type~I bursts. Since all of the
four events occurred in the FB, we conclude that they must be flares, of which the light curves 
happen to resemble those of X-ray bursts (such as b1, b3 and b5). We will not discuss 
these four events further in the paper and denote the remainder of the ten events 
as bursts, since we will show below that they are genuine type~I X-ray bursts.

\begin{figure*}
 \resizebox{12cm}{!}{\includegraphics[angle=-90, clip, bb=28 41 574 744]{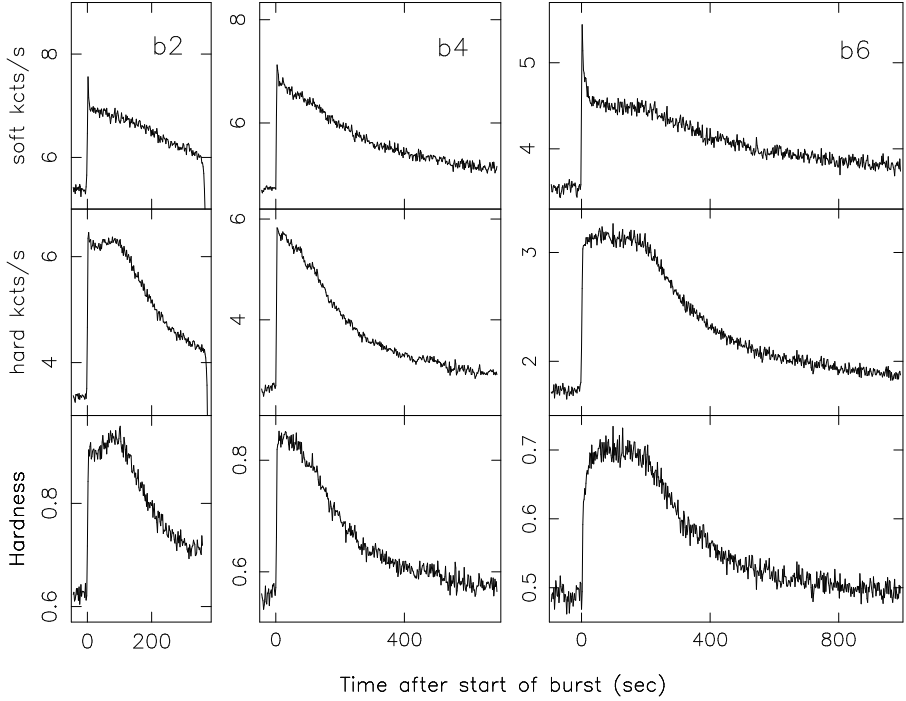}}
   \caption{Same as Fig.~\ref{plot_b1_b3_b5}, but at a time resolution of 2\,s,
for the long bursts b2, b4 and b6.
Note that b2 was interrupted by a South Atlantic Anomaly (SAA) passage as can be seen by the sudden decrease in
count rate.
The low and high energy ranges are 
1.9--6.2\,keV and 6.2--$\simeq$60\,keV, respectively, for b2, 
1.9--6.2\,keV and 6.2--19.6\,keV, respectively, for b4, 
and 2.1--7.1\,keV and 7.1--19.9\,keV, respectively, for b6.}
   \label{plot_b2_b4_b6}
\end{figure*}

\begin{figure*}
 \resizebox{12cm}{!}{\includegraphics[angle=-90, clip, bb=28 41 574 761]{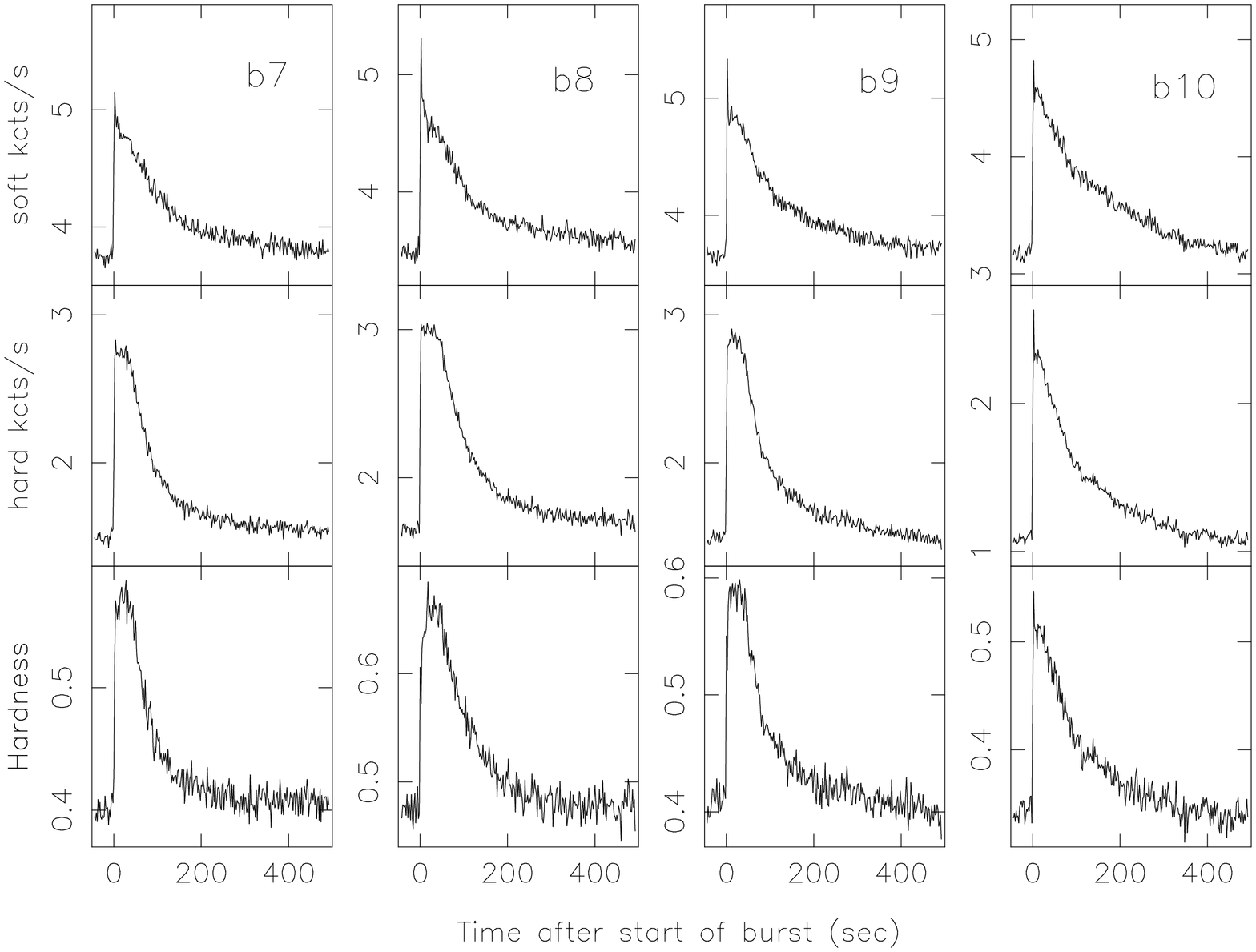}}
   \caption{Same as Fig.~\ref{plot_b2_b4_b6}, for the last four long bursts observed in 
1999 Oct (b7--b10). The low and high energy ranges are 
1.9--7.1\,keV and 7.1--19.9\,keV, respectively.}
   \label{plot_b7-b10}
\end{figure*}

In Fig.~\ref{plot_b1_b3_b5} we show the light curves
of the three short bursts b1, b3 and b5, at low and high energies, 
with the corresponding hardness curves, all at a time resolution of 0.125\,s.
All three bursts show a fast rise (typically less than 0.5\,s) and an exponential decay with
a decay time of $\simeq$2\,s (see Table~2). During the rise the emission hardens; as the bursts 
decay, the emission becomes softer. The main difference between the three
bursts is the fact that the peak of burst b3 is about 25\%\ lower than the
other two bursts; it looks like a `failed' burst. It also has two peaks, as if some 
new unstable burning occurred, $\simeq$5\,s after the start of the burst.

\begin{figure}
 \resizebox{\hsize}{!}{\includegraphics[angle=-90, clip, bb=23 41 574 590]{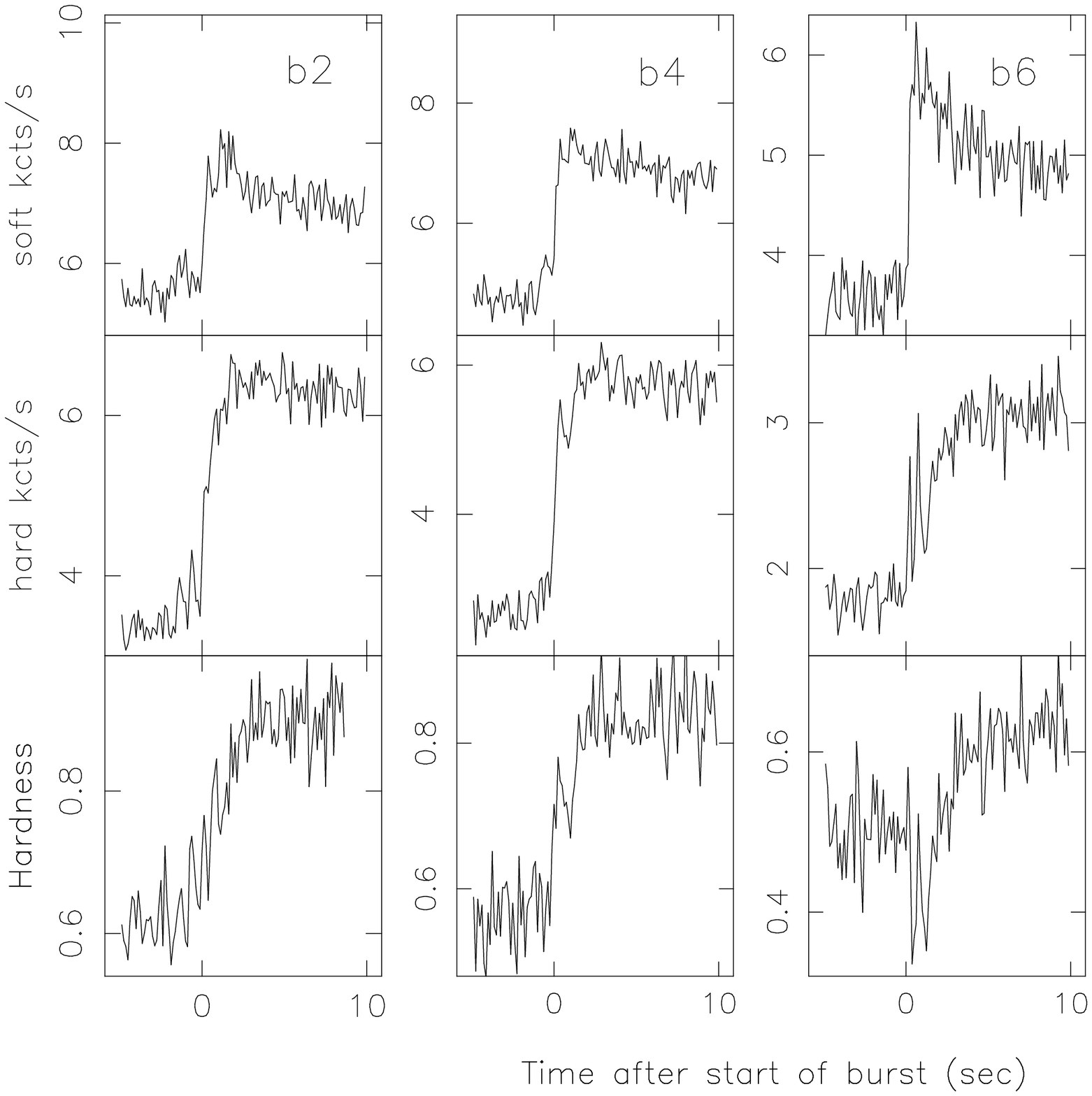}}
   \caption{Same as Fig.~\ref{plot_b2_b4_b6}, but 5\,s before and 10\,s after the start of 
the bursts b2, b4 and b6, at a time resolution of 0.125\,s.}
   \label{plot_start_b2_b4_b6}
\end{figure}

\begin{figure*}
 \resizebox{12cm}{!}{\includegraphics[angle=-90, clip, bb=28 41 574 771]{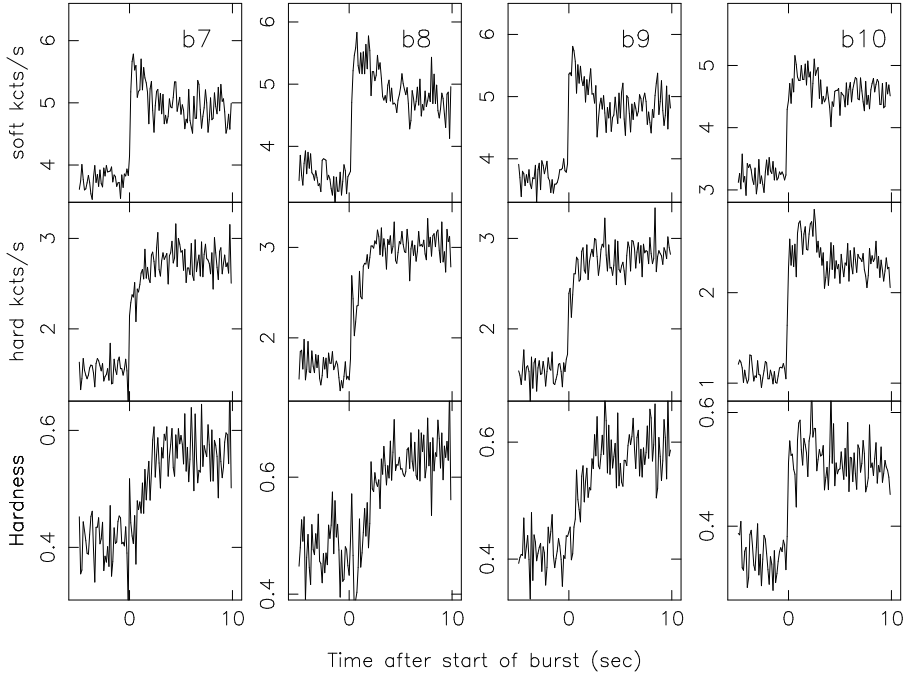}}
   \caption{Same as Fig.~\ref{plot_b7-b10}, but 5\,s before and 10\,s after the start of 
the last four bursts observed in 1999 Oct (b7--b10),
at a time resolution of 0.125\,s.}
   \label{plot_start_b7-b10}
\end{figure*}

In Figs.~\ref{plot_b2_b4_b6} and \ref{plot_b7-b10} we show the light curves of the long bursts 
b2, b4, and b6--b10, at low and high energies, with the corresponding hardness curves, all
at a time resolution of 2\,s.
In Figs.~\ref{plot_start_b2_b4_b6} and \ref{plot_start_b7-b10} we focus on the start of these bursts, 
all at a time resolution of 0.125\,s.
Again the rise times are very short (also typically less than 0.5\,s), but the decay times are
much longer, with decay times in the range $\simeq$70--280\,s (see Table~2). 
Apart from the fact that the hard burst emission decays faster than the soft burst
emission (i.e.\ spectral softening), there are more pronounced differences between
the light curves in the two energy bands. In Figs.~\ref{plot_b2_b4_b6} and \ref{plot_b7-b10} 
one sees that all 
the low energy light curves show a kind of
spike at the start of the decay.
These spikes last for a few seconds in most cases; however, for burst b6 it seems to last
$\simeq$15\,s (with an exponential decay time of 9$\pm$2\,s). 
At high energies no such spikes occur (except for burst b10); instead the bursts have more flat-topped
peaks, with durations ranging from tens of seconds to $\simeq$200\,s. Burst b6 is
the nicest example of this.
Zooming in on the start of these long bursts, it becomes clear that the rise is somewhat slower
at high energies with respect to low energies (causing the hardening of the emission during
the early phase of the burst). Also, in bursts b4 and b6--b9 
very short ($<$0.5\,s) spikes occur during the rise
in the high energy light curve, which again are especially evident in burst b6 (two spikes!). 
This causes the hardness to drop on the same time scale in some of these bursts.
Later we will show that this corresponds to fast radius expansion/contraction episodes.

\subsection{Spectral behaviour}

\subsubsection{Persistent emission before the bursts}

\begin{table*}
\caption{Persistent emission spectral parameters$^a$}
\begin{tabular}{ccccccccccc}
\hline
\multicolumn{1}{c}{$\#$} & \multicolumn{1}{c}{F$_{\rm pers}$$^b$} &
\multicolumn{1}{c}{kT$_{\rm bb}$} &  \multicolumn{1}{c}{R$_{\rm bb,10}$$^c$} &
\multicolumn{1}{c}{$\Gamma$$^d$} & \multicolumn{1}{c}{E$_{\rm cut}$} &
\multicolumn{1}{c}{norm.$^e$} &  \multicolumn{1}{c}{gnorm.$^f$} & 
\multicolumn{1}{c}{$\chi^2_{\rm red}$/} & \multicolumn{1}{c}{bb$^g$} & 
\multicolumn{1}{c}{P$^h$} \\
\multicolumn{1}{c}{~} & \multicolumn{1}{c}{(2--20\,keV)} &
\multicolumn{1}{c}{(keV)} &  \multicolumn{1}{c}{(km)} &
\multicolumn{1}{c}{~} & \multicolumn{1}{c}{(keV)} &
\multicolumn{1}{c}{~} & \multicolumn{1}{c}{~} & 
\multicolumn{1}{c}{dof} & \multicolumn{1}{c}{~} & 
\multicolumn{1}{c}{~} \\
\hline
 b1 & 2.0$\pm$0.3 & 1.15$\pm$0.06 & 14.1$\pm$1.5 & 1.0$\pm$0.1 & 4.5$\pm$0.2 & 3.8$\pm$0.4 & 0.013$\pm$0.002 & 0.89/47 & 16 & 5$\times$10$^{-10}$ \\
 b2 & 2.2$\pm$0.5 & 1.10$\pm$0.03 & 22.1$\pm$2.2 & 0.4$\pm$0.2 & 3.8$\pm$0.2 & 1.5$\pm$0.4 & 0.008$\pm$0.002 & 0.56/40 & 29 & 4$\times$10$^{-18}$ \\
 b3 & 1.9$\pm$0.3 & 1.13$\pm$0.03 & 19.2$\pm$1.5 & 0.9$\pm$0.1 & 4.2$\pm$0.2 & 2.6$\pm$0.4 & 0.013$\pm$0.002 & 0.58/40 & 29 & 1$\times$10$^{-18}$ \\
 b4 & 2.0$\pm$0.3 & 1.10$\pm$0.04 & 20.3$\pm$1.8 & 0.7$\pm$0.1 & 4.0$\pm$0.2 & 2.3$\pm$0.4 & 0.014$\pm$0.002 & 0.85/40 & 28 & 6$\times$10$^{-15}$ \\
 b5 & 2.3$\pm$0.3 & 1.16$\pm$0.05 & 17.5$\pm$1.7 & 0.9$\pm$0.1 & 4.5$\pm$0.2 & 3.2$\pm$0.4 & 0.010$\pm$0.002 & 0.59/40 & 23 & 5$\times$10$^{-15}$ \\
 b6 & 2.5$\pm$0.1 & --- & --- & 1.03$\pm$0.03 & 5.1$\pm$0.1 & 4.8$\pm$0.1 & 0.012$\pm$0.002 & 0.71/35 & $<$7$^i$ & 8 \\
 b7 & 2.5$\pm$0.1 & --- & --- & 1.17$\pm$0.03 & 4.6$\pm$0.1 & 7.0$\pm$0.2 & 0.015$\pm$0.002 & 1.22/35 & $<$2$^i$ & 30 \\
 b8 & 2.4$\pm$0.5 & 1.08$\pm$0.13 & 11.0$\pm$4.3 & 1.0$\pm$0.1 & 5.3$\pm$0.3 & 3.8$\pm$0.5 & 0.011$\pm$0.002 & 1.13/33 & 6 & 1 \\
 b9 & 2.4$\pm$0.4 & 1.26$\pm$0.08 & 11.6$\pm$1.5 & 1.0$\pm$0.1 & 4.9$\pm$0.3 & 4.3$\pm$0.4 & 0.010$\pm$0.003 & 0.87/33 & 13 & 8$\times$10$^{-5}$ \\ 
b10 & 1.9$\pm$0.3 & 1.18$\pm$0.05 & 15.1$\pm$1.4 & 1.1$\pm$0.1 & 4.5$\pm$0.3 & 3.7$\pm$0.5 & 0.011$\pm$0.002 & 1.16/33 & 22 & 5$\times$10$^{-7}$ \\
\hline
\multicolumn{11}{l}{\footnotesize $^a$\,N$_{\rm H}$ was fixed to 2$\times$10$^{22}$\,atoms\,cm$^{-2}$, see text.} \\
\multicolumn{11}{l}{\footnotesize $^b$\,Unabsorbed persistent flux in 10$^{-8}$\,erg\,s$^{-1}$\,cm$^{-2}$.}\\
\multicolumn{11}{l}{\footnotesize $^c$\,Apparent black-body radius at 10\,kpc.}\\
\multicolumn{11}{l}{\footnotesize $^d$\,Power-law photon index.}\\
\multicolumn{11}{l}{\footnotesize $^e$\,Power-law normalization in photons\,keV$^{-1}$\,cm$^{-2}$\,s$^{-1}$ at 1\,keV.}\\
\multicolumn{11}{l}{\footnotesize $^f$\,Normalization in photons\,cm$^{-2}$\,s$^{-1}$ in the line
modeled by a Gaussian fixed at 6.7\,keV and width ($\sigma$) of 0.1\,keV.}\\
\multicolumn{11}{l}{\footnotesize $^g$\,Black-body emission contribution to the total persistent flux in \%\ (2--20\,keV).} \\ 
\multicolumn{11}{l}{\footnotesize $^h$\,Probability (in\%\/) that inclusion of a black-body component is {\it not} significant.} \\ 
\multicolumn{11}{l}{\footnotesize $^i$\,95\%\ confidence upper limit, using kT$_{\rm bb}$=1.14\,keV, see text.} \\ 
\end{tabular}
\end{table*}

The persistent emission just before the bursts b6 and b7 can be satisfactorily 
($\chi^2_{\rm red}$=0.7 and 1.2, respectively, for 35 degrees of freedom, dof) described by 
an absorbed cut-off power-law component plus a Gaussian line. 
For the persistent 
emission before the remainder of the bursts this is not a satisfactory model 
($\chi^2_{\rm red}$ ranges from 1.4 with 35 dof for b8 to 5.5 with 42 dof for b3). 
An additional component is necessary to improve the fits. 
We used the F-test to calculate whether the additional component was indeed significant.
For the additional component we chose a black body, as is generally 
used when modeling the X-ray spectra of 
bright LMXBs (e.g.\ White et al.\ 1986; Christian \&\ Swank 1997; Church \&\ Ba\l uci\'nska-Church 2001). 
Note that more 
complicated models are necessary to describe the spectra of GX\,17+2 with better energy resolution 
and broader energy range (e.g.\ Di Salvo et al.\ 2000). However, our approach is adequate for this paper.
In the rightmost column of Table 3 we give the results of our F-tests, i.e.\ the probability that
the addition of the black-body component in the spectral fits is {\it not} significant. 
We conclude that except for bursts b6 and b7, an additional component is warranted on the $\gtrsim$99\%\
level. In Table~3 we present the results of the spectral fits to the persistent emission. 
Some of the $\chi^2_{\rm red}$ values are rather small (bursts b2, b3, b5), perhaps due to a slight
overestimate of the systematic error added (1\%\/).
The unabsorbed persistent flux just before the bursts during the various observations varied between 
1.9--2.5$\times$10$^{-8}$\,erg\,s$^{-1}$\,cm$^{-2}$ (2--20\,keV). 
For bursts b6 and b7 we determined single parameter 95\%\ confidence upper limits 
(using $\Delta\chi^2$=2.71) on the black-body contribution, by including a black-body component
in the spectral fits, and fixing the temperature to its mean value
derived for the persistent spectra of the other bursts, i.e.\ kT$_{\rm bb}$=1.14\,keV.
The black-body component
contribution to the persistent emission varied between less than 2\%\ 
(burst b7) up to 29\%\ (bursts b2, b3) in the 2--20\,keV band (Table 3).

\begin{figure*}
 \resizebox{12cm}{!}{\includegraphics[angle=-90, clip, bb=50 75 563 769]{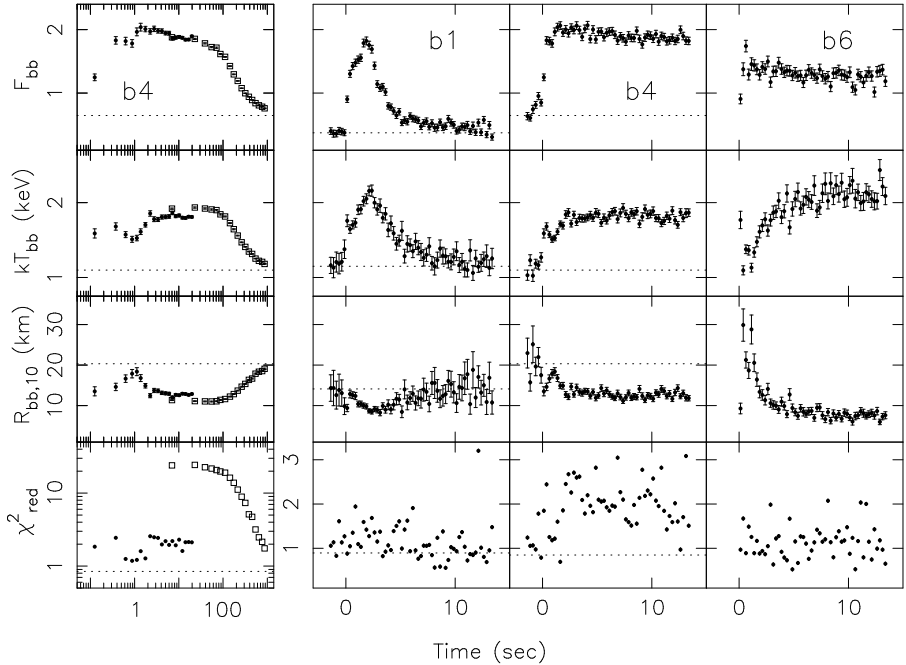}}
   \caption{{\it Leftmost panel}: Two-component spectral fit results for the total burst emission of 
burst b4 plotted on a logarithmic time scale.
The filled circles and open squares represent the fit results of the 0.25\,s and 16\,s spectra, 
respectively. The data have been logarithmically rebinned in time for clarity.
The values for the persistent black-body component have been indicated by a dotted line.
{\it Right panels}: Two-component spectral fit results for the 0.25\,s spectra of the total burst emission of 
burst b1, b4 and b6, see text.
The values for the persistent black-body component have been indicated by a dotted line
for burst b1 and b4.
{\it Both panels}: from top to bottom: bolometric black-body flux, F$_{\rm bb}$, 
in 10$^{-8}$\,erg\,s$^{-1}$\,cm$^{-2}$,
black-body temperature, kT$_{\rm bb}$, apparent black-body radius, R$_{\rm bb,10}$, at 10\,kpc, 
and goodness of fit expressed in $\chi^2_{\rm red}$. For bursts b1, b4 and b6 the number 
of dof is 22, 18 and 16, respectively, for the 0.25\,s spectral fits. 
For burst b4 the number of dof is 44 for the 16\,s spectral fits. 
Note the difference in scales of $\chi^2_{\rm red}$ in the leftmost panel with respect to the
other panels.}
   \label{plot_pars_b4log_b1_b4_b6}
\end{figure*}

\begin{figure*}
\centerline{
 \resizebox{8cm}{!}{\includegraphics[angle=-90, clip, bb=112 18 571 550]{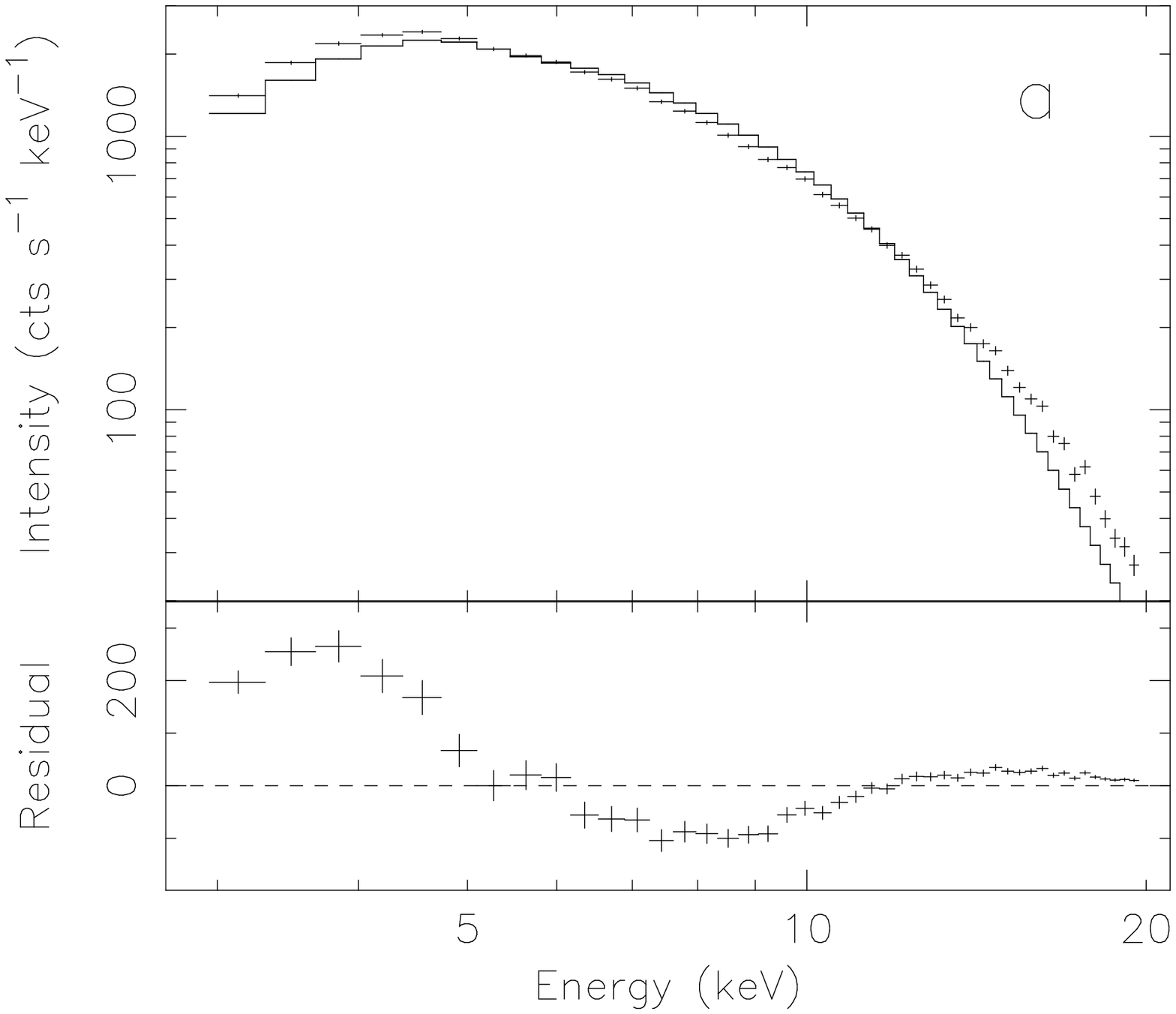}}
 \resizebox{8cm}{!}{\includegraphics[angle=-90, clip, bb=112 18 571 550]{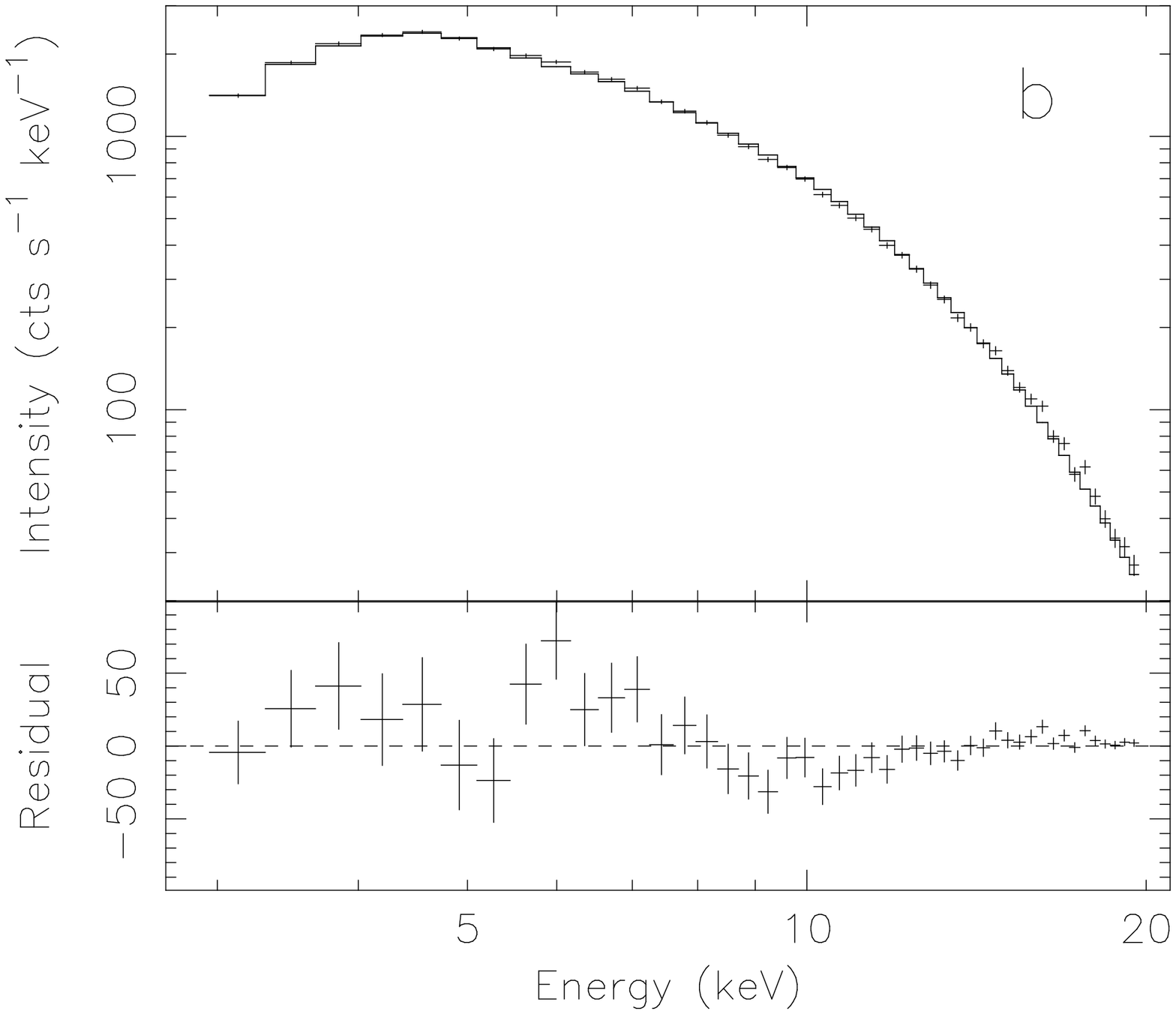}}
}
\caption{{\it a}: At the top the first 16\,s 
spectrum observed after the start of burst b2 is displayed.
A two-component fit is shown, i.e.\ a single black-body and cut-off power-law 
plus Gaussian line (subjected to interstellar
absorption). The parameters of the cut-off power-law component plus Gaussian line 
are fixed to the values derived for the persistent emission. 
At the bottom the residuals after subtracting the best model from the observed spectrum is displayed.
The fit is clearly bad ($\chi^2_{\rm red}$/dof=24.2/44).
{\it b}: At the top the same first 16\,s spectrum at the beginning of burst b2 is displayed.
Now a three-component fit is shown, i.e.\ two black-body components and one cut-off power-law
component (subjected to interstellar absorption). The cut-off power-law and one black-body component
parameters are fixed to the values derived for the persistent emission.
At the bottom the residuals after subtracting the best model from the observed spectrum is displayed.
The fit has clearly improved ($\chi^2_{\rm red}$/dof=1.8/44).}
\label{burstb2_b3_bb}
\end{figure*}

\subsubsection{Persistent emission during the bursts?}

\paragraph{3.2.2.1 Previous EXOSAT results} ~\\

\noindent
Usually it is assumed that the persistent emission is not influenced by the burst and that
one can, therefore, study the burst by subtracting the persistent emission
from the total source emission.
This is referred to as the `standard' burst spectral analysis (see e.g.\ Sztajno et al.\ 1986).
However, if the neutron star photosphere contributes significantly to the
persistent emission, this approach is not correct, if the burst emission
originates from the same region (van Paradijs \&\ Lewin 1986). 
In this case the spectral fits to the net burst spectra yield a systematically larger 
black-body temperature, T$_{\rm bb}$, 
and smaller apparent black-body radius, R$_{\rm bb}$, especially near the end of the burst when the
net burst flux is low. 
In fact, in this case 
of incorrect subtraction of the persistent emission, 
the net burst spectrum is not a black-body.
Van Paradijs \&\ Lewin (1986) proposed to fit the total source spectrum 
with a two-component model, a black-body component and a non-black-body component.
During the burst the non black-body component is fixed to what is found in the persistent emission, and 
the black-body component is left free.
The black-body component should include all emission from the neutron star photosphere.
The underlying idea is that the non black-body component arises from the accretion process, and is 
not influenced by the X-ray burst.

GX\,17+2 is a bright X-ray source, and presumably the neutron star contributes significantly to the 
persistent emission (Van Paradijs \&\ Lewin 1986; Sztajno et al.\ 1986). 
Since the source is bright, compared to most other burst sources 
the net flux at the peak of the burst relative to the persistent
flux is rather low. Sztajno et al.\ (1986) found that
the black-body component contributed $\simeq$40\%\ 
to the persistent emission just before the two bursts observed by EXOSAT. 
Using the `standard' burst spectral analysis Sztajno et al.\ (1986) found relatively 
high black-body temperatures (kT$_{\rm bb}$$\simeq$2--3\,keV) and relatively small 
apparent black body radii at a distance of 10\,kpc (R$_{\rm bb,10}$$\simeq$3--5\,km, 
for isotropic emission) for the short ($\simeq$10\,s) burst. For the long ($>$5\,min) burst,
kT$_{\rm bb}$ only showed a small change from $\simeq$2.1\,keV at the peak of the burst
to $\simeq$1.7\,keV near the end of the burst, with R$_{\rm bb,10}$ decreasing  
from $\simeq$7\,km to $\simeq$4\,km. Their fits were, however, satisfactory, 
with $\chi^2_{\rm red}$ of 0.6--1.2. 
Using the two-component model, Sztajno et al.\ (1986) found that
during the short burst the values for T$_{\rm bb}$ were in the range normally seen in type~I X-ray bursts, 
and that the systematic decrease in R$_{\rm bb}$ had disappeared.
The $\chi^2_{\rm red}$ values for the two-component model fits ranged 
between 0.6 and 1.3, except for one fit, for which
$\chi^2_{\rm red}$=1.6, i.e.\ slightly worse than the `standard' spectral analysis. 
They attributed this to the smaller relative error
in the total burst data than in the net-burst data, for which the persistent emission was subtracted.
Note that in this case Sztajno et al.\ (1986) found a slight increase in R$_{\rm bb,10}$, 
apparently anti-correlated with T$_{\rm bb}$, which they argue was 
due to the non-Planckian shape of the spectrum of a hot neutron star
(van Paradijs 1982; see Titarchuk 1994; Madej 1997, and references therein; see, however, Sect.~5.2).

\paragraph{3.2.2.2 Our RXTE results} ~\\

\noindent
Guided by the results of Sztajno et al.\ (1986) discussed in Sect.~3.2.2.1, 
we first fitted the total burst data 
using a black body and a cut-off power law plus a Gaussian line.
The parameters of the absorbed cut-off power law and
Gaussian line were fixed to the values found for the persistent emission before the burst (Table~3). 
Using this model we obtained good fits 
to the 16\,s spectra 
of bursts b6 and b7, for which the persistent emission did not contain a significant 
black-body contribution
($\chi^2_{\rm red}$ of $\simeq$1 for 37 dof).
However, the 16\,s spectral fits were bad whenever the
persistent emission spectra contained a black-body component;
the fits became worse as the persistent black-body
contribution became stronger (for the first $\sim$100\,s of the burst:
$\chi^2_{\rm red}$/dof$\simeq$1.5--3/37 [burst b8], 
$\chi^2_{\rm red}$/dof$\simeq$2.5--8/37 [burst b9], 
$\chi^2_{\rm red}$/dof$\simeq$5--10/37 [burst b10], 
$\chi^2_{\rm red}$/dof$\simeq$20--24/44 [burst b4], 
$\chi^2_{\rm red}$/dof$\simeq$22--25/44 [burst b2]).
The worst $\chi^2_{\rm red}$ occurred near the 
peak of the bursts. For instance, Fig.~\ref{plot_pars_b4log_b1_b4_b6} shows
the best-fit parameters and $\chi^2_{\rm red}$ for burst b4, 
for which the persistent emission had a 
black-body contribution of $\simeq$28\%\ (2--20\,keV). 
An example of a burst spectrum and the result of the two-component fit 
is shown in Fig.~\ref{burstb2_b3_bb}a for burst b2. 
The $\chi^2_{\rm red}$ with 44 dof for burst b4
decreases from $\simeq$24 at the beginning of the burst to $\simeq$2 near the end of the burst.
A similar trend is seen for the 0.25\,s burst spectral fits: 
$\chi^2_{\rm red}$ of $\simeq$2 for 18 dof during the first $\simeq$10\,s of burst b4 
and an $\chi^2_{\rm red}$ of $\simeq$1 
for 16 dof during the first $\simeq$10\,s of burst b6 (see 
Fig.~\ref{plot_pars_b4log_b1_b4_b6}). 
The $\chi^2_{\rm red}$ values for 0.25\,s spectral fits were much lower than those for the
16\,s spectral fits, 
due to the much lower signal to noise and lower spectral resolution of the 0.25\,s spectra.
Note that the high values of $\chi^2_{\rm red}$ for the 16\,s spectra are not due to
fast spectral variations within the time the spectra are accumulated. This only applies
to spectra which include the first few seconds of the bursts, 
during the short radius expansion and initial contraction phase (see Sect.~3.2.4).
Halfway the decline of burst b4, R$_{\rm bb}$ seems to increase
again; this is also seen in the other long bursts (except burst b6), and in the short bursts
(e.g.\ burst b1, see Fig.~\ref{plot_pars_b4log_b1_b4_b6}). 
This is similar to that seen by Sztajno et al.\ (1986), but their fits 
seemed better (see Sect.~3.2.2). 
This is (probably) due to the much lower 
signal to noise ratio of the EXOSAT/ME data.

\begin{figure}
 \resizebox{\hsize}{!}{\includegraphics[angle=-90, clip, bb=48 43 563 550]{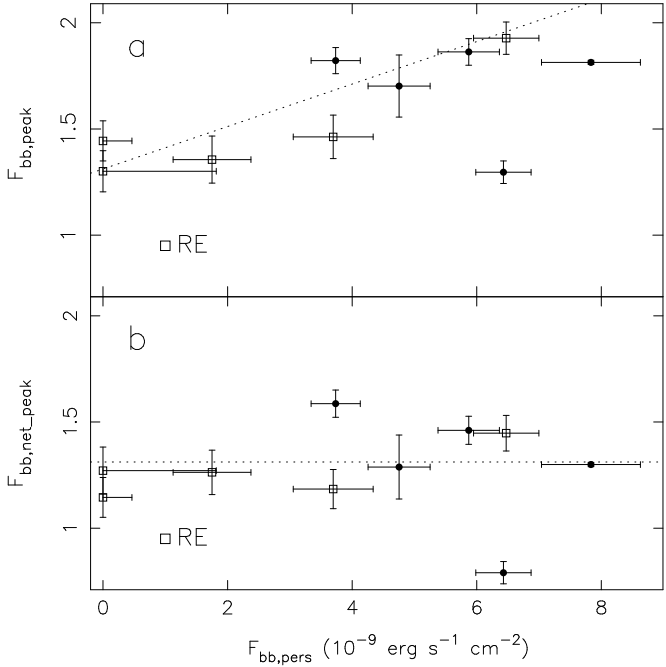}}
\caption{{\it a}: Bolometric peak black-body flux (see text), F$_{\rm bb,peak}$, 
in units of 10$^{-8}$\,erg\,s$^{-1}$\,cm$^{-2}$ versus the bolometric 
persistent black-body flux, F$_{\rm bb,pers}$. Radius expansion (RE) bursts are denoted with an open square.
The dotted line represents a fit to the data points (excluding burst b2 and b3) to the function
F$_{\rm bb,peak} = {\rm F}_{\rm bb,pers} + C$. The constant $C$ corresponds to 
(1.31$\pm$0.03)$\times$10$^{-8}$\,erg\,s$^{-1}$\,cm$^{-2}$.
{\it b}: 
Bolometric peak black body flux from spectral fits after subtraction of the persistent emission (see text),
F$_{\rm bb,net\_peak}$, in units of 10$^{-8}$\,erg\,s$^{-1}$\,cm$^{-2}$ versus 
F$_{\rm bb,pers}$. Radius expansion (RE) bursts are denoted with an open square.
The dotted line represents the constant $C$.}
\label{plot_peak_pers}
\end{figure}

The mean bolometric black-body flux during the first 10\,s after 
the peak, F$_{\rm bb,peak}$, of bursts b4 
is higher than that of burst b6.
The difference is 0.63($\pm$0.12)$\times$10$^{-8}$\,erg\,s$^{-1}$\,cm$^{-2}$
(for the uncertainty in F$_{\rm bb,peak}$ we used the rms variation in the bolometric black-body fluxes).
This is consistent with the bolometric black body contribution in the persistent emission,
F$_{\rm bb,pers}$, before burst b4 (indicated by the dotted lines in the top panels for burst b4 
in Fig.~\ref{plot_pars_b4log_b1_b4_b6}): 
F$_{\rm bb,pers}$=0.65($\pm$0.05)$\times$10$^{-8}$\,erg\,s$^{-1}$\,cm$^{-2}$.
We measured F$_{\rm bb,peak}$ and F$_{\rm bb,pers}$ for all the bursts to see if this 
effect is seen in other bursts as well. We used the highest observed black body flux, 
except for burst b2 where we used the first 16\,s measurement
after the start of the burst. The result is displayed in Fig.~\ref{plot_peak_pers}a.
Indeed, F$_{\rm bb,peak}$ differs between the bursts and is 
clearly correlated with F$_{\rm bb,pers}$ (except for burst b3). 
Such a relation can be expected if F$_{\rm bb,peak}$ is close to a certain upper limit (presumably
the Eddington limit) and includes a (certain fraction) of the persistent emission.
We can reasonably fit the data points to 
the function F$_{\rm bb,peak} = {\rm F}_{\rm bb,pers} + C$ 
(excluding bursts b2 and b3; burst b2 did not cover the first 10\,s after the peak of the burst, and
burst b3 was weaker than the other bursts). The resulting $\chi^2_{\rm red}$=2.0 for 7 dof.
We find $C=(1.31\pm 0.03)$$\times$10$^{-8}$\,erg\,s$^{-1}$\,cm$^{-2}$; this is close to the values
of F$_{\rm bb,peak}$ for bursts b6 and b7 which had no significant black-body contribution in the 
preceding persistent emission. 
In Fig.~\ref{plot_peak_pers} 
we have also indicated whether a burst was a radius expansion event
(see Sect.~3.2.4). It can be seen in Fig.~\ref{plot_peak_pers}a
that the bursts with no radius expansion/contraction phase (except burst b3)
have somewhat higher values of F$_{\rm bb,peak}$ than the radius expansion bursts
(except burst b4).
Theoretically, during the radius expansion/contraction 
phases the flux should equal the Eddington limit. Therefore, the (peak) fluxes observed 
for different radius expansion bursts should have similar values, whereas for
bursts with no radius expansion/contraction phase they should be either similar or smaller.
This is not the case.

\begin{figure}
 \resizebox{\hsize}{!}{\includegraphics[angle=-90, clip, bb=50 75 563 518]{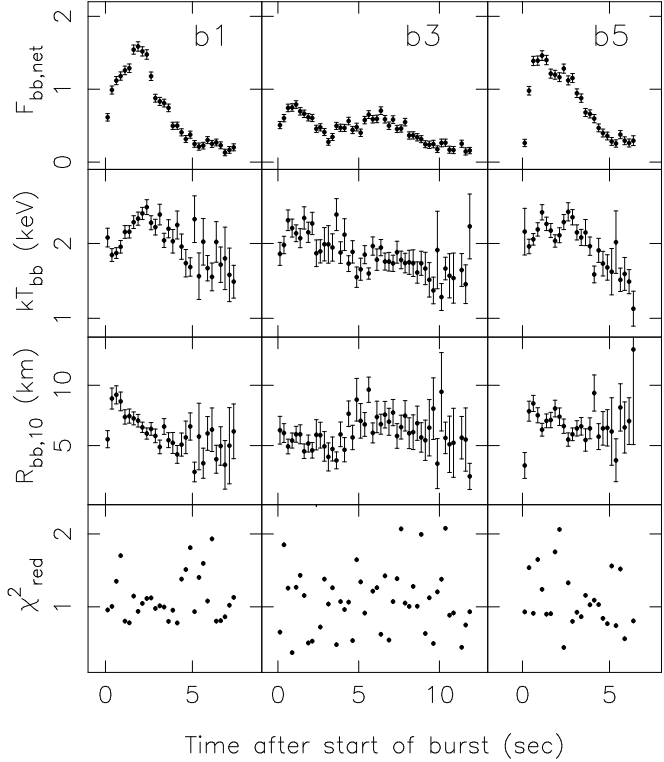}}
   \caption{Spectral fit results for the net burst emission of burst b1, b3 and b5; from top to bottom:
bolometric black-body flux, F$_{\rm bb,net}$, in 10$^{-8}$\,erg\,s$^{-1}$\,cm$^{-2}$,
black-body temperature, kT$_{\rm bb}$, apparent black-body radius, R$_{\rm bb,10}$, at 10\,kpc, 
and goodness of fit expressed in reduced $\chi^2$. The number of dof is 22 for burst b1 and
18 for bursts b3 and b5.}
   \label{plot_pars_b1_b3_b5_bb}
\end{figure}

To summarize, we find that the method proposed by Van Paradijs \&\ Lewin (1986) and applied by
Sztajno et al.\ (1986) to the bursts observed by EXOSAT does not work for our 
bursts observed with the RXTE/PCA. The two-component spectral fits during the bursts give bad results 
whenever there is a black-body contribution to the per-burst 
persistent emission. Moreover, the total burst 
peak fluxes, F$_{\rm bb,peak}$ are different from burst to burst 
and depend on the amount of the black-body contribution 
to the persistent emission, F$_{\rm bb,pers}$. This is not what one would expect if during some of the bursts
a limit is reached (presumably the Eddington limit).
We conclude that the persistent black-body emission does {\em not} disappear
during the burst.
An example is displayed in Fig.~\ref{burstb2_b3_bb}b. We show again the first 16\,s 
spectrum during burst b2 plus now the results of a three-component fit, i.e.\ two black bodies and a
cut-off power-law plus Gaussian line (subjected to interstellar absorption). The parameters of the 
cut-off power law plus Gaussian line and one black-body component have been fixed to those found for the 
persistent emission (so the fit does have the same number of dof
as the one displayed in Fig.~\ref{burstb2_b3_bb}a).
Clearly, the fit has improved considerably.
This means that the burst emission is decoupled from the persistent emission, and  
therefore the `standard'
spectral analysis should work in this case. This is the subject of the next subsection.

\subsubsection{`Standard' spectral analysis: net burst emission}

\begin{figure*}
 \resizebox{\hsize}{!}{\includegraphics[angle=-90, clip, bb=50 37 563 769]{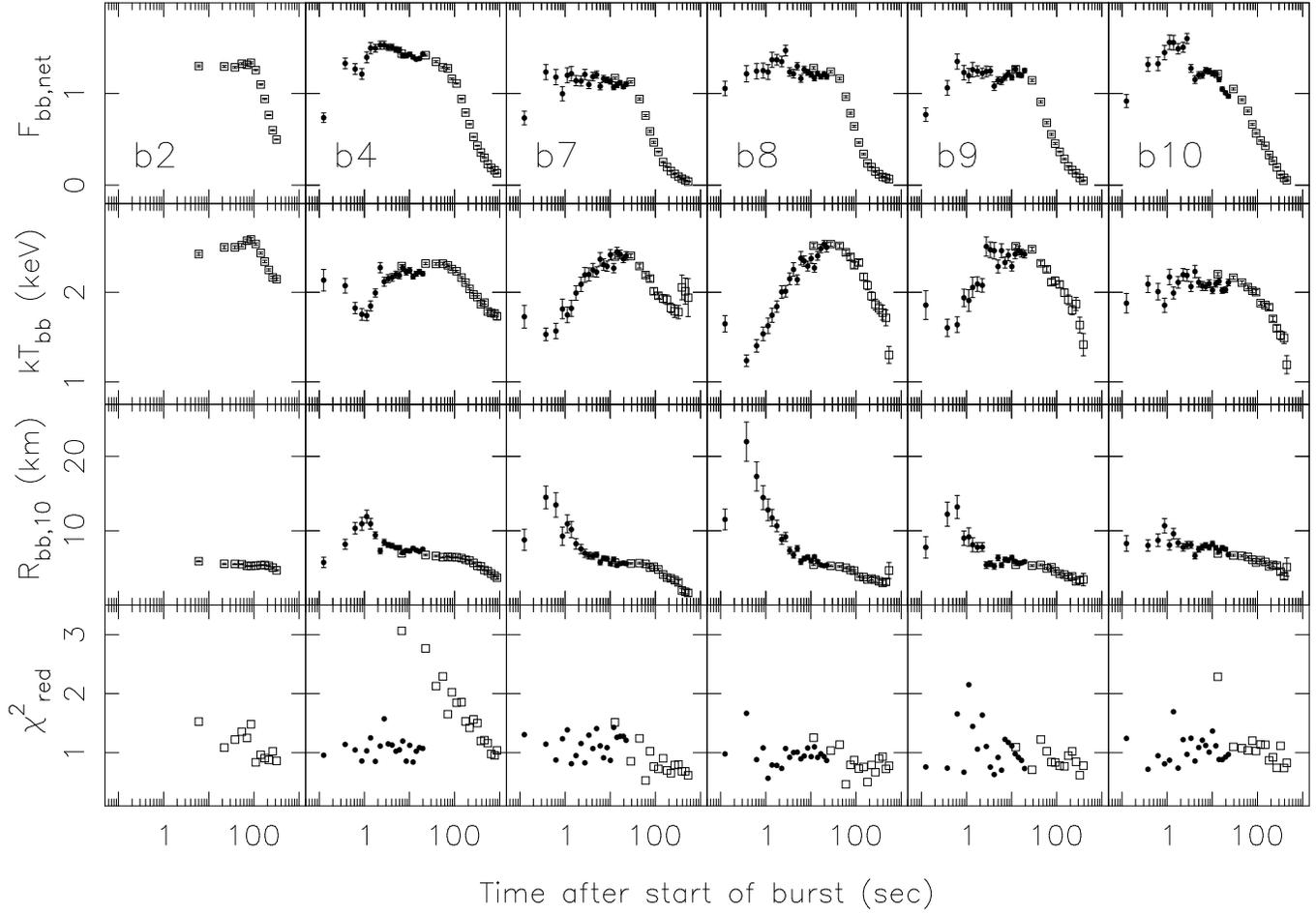}}
\caption{Spectral fit results for the net burst emission of 
bursts b2, b4 and b7--b10 plotted on a logarithmic time scale; 
from top to bottom: bolometric black-body flux, F$_{\rm bb,net}$, in 10$^{-8}$\,erg\,s$^{-1}$\,cm$^{-2}$,
black-body temperature, kT$_{\rm bb}$, apparent black-body radius, R$_{\rm bb,10}$, at 10\,kpc, 
and goodness of fit expressed in reduced $\chi^2$.
The filled circles and open squares represent the fit results of the 0.25\,s and 16\,s spectra, 
respectively. The data have been logarithmically rebinned for clarity.
For the 0.25\,s spectral fits the number of dof is 18 for burst b4 and 16 for bursts b7--b10.
For the 16\,s spectral fits the number of dof is 44 for bursts b2 and b4, and 37 for bursts b7--b10.}
\label{plot_pars_long_b2_b4_b7-b10_bb_log}
\end{figure*}

\begin{figure*}
 \resizebox{\hsize}{!}{\includegraphics[angle=-90, clip, bb=50 37 566 769]{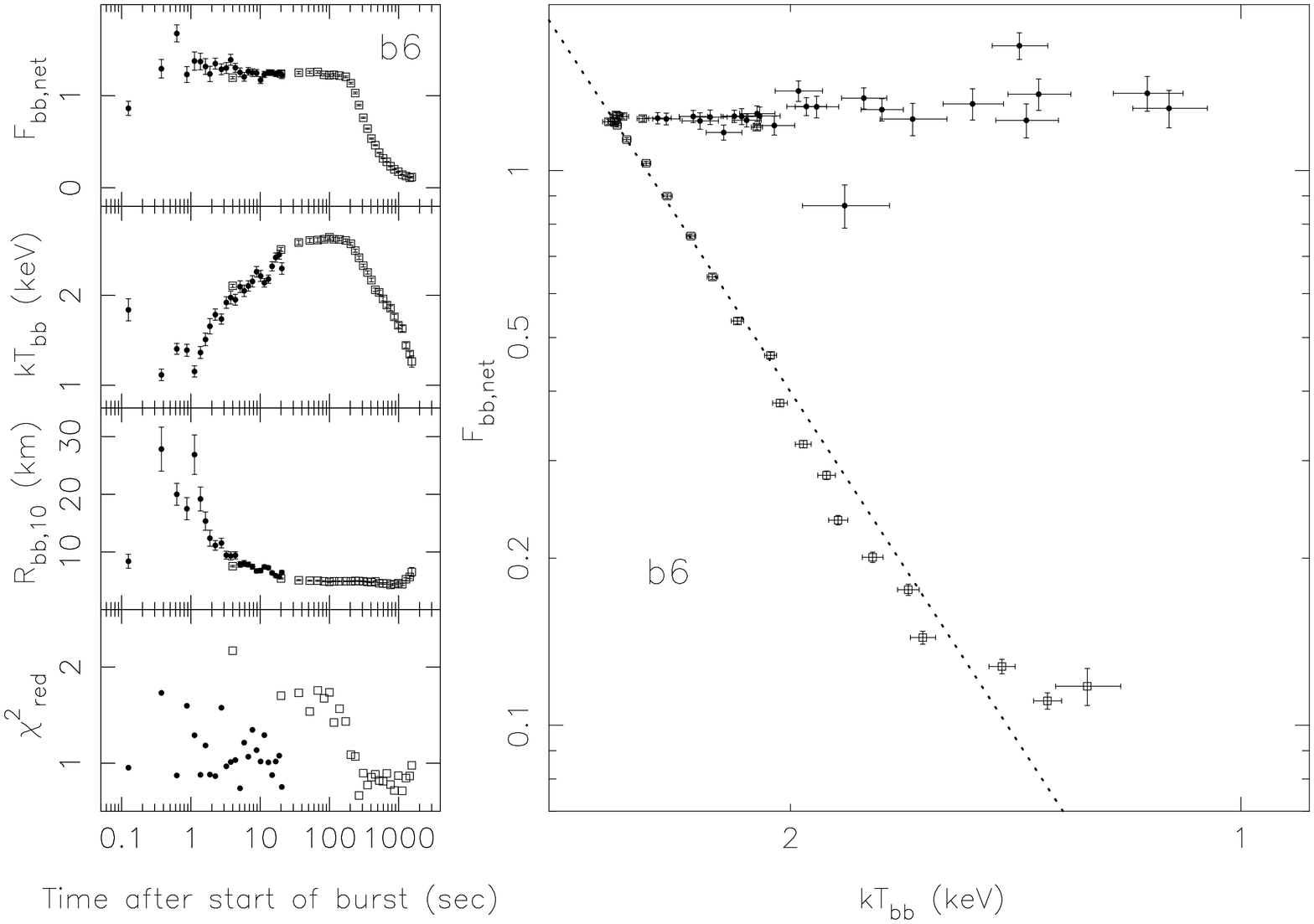}}
\caption{{\it Left panel}: same as Fig.~\ref{plot_pars_long_b2_b4_b7-b10_bb_log}, but for burst b6, the longest
burst in our sample. {\it Right panel}: Bolometric black-body flux (F$_{\rm bb,net}$ in 
10$^{-8}$\,erg\,s$^{-1}$\,cm$^{-2}$) versus black-body temperature (kT$_{\rm bb}$) for burst b6. 
The dotted line represents the fit to the cooling track of the burst for data points with 
F$_{\rm bb,net}>4\times 10^{-9}$\,erg\,s$^{-1}$\,cm$^{-2}$, see text. Note
that kT$_{\rm bb}$ increases from right to left.
For the 0.25\,s and 16\,s spectral fits the number of dof is 16 and 37, respectively.}
\label{plot_pars_b6_bb_all_plus_fbolvst}
\end{figure*}

When modeling the net burst spectra (i.e.\ spectra obtained after subtraction of the 
pre-burst persistent 
emission from the total source spectrum) by a black body,
the spectral fits are almost all satisfactory ($\chi^2_{\rm red}$ values of $\simeq$1).
The fit results are shown for the 3 short bursts 
in Fig.~\ref{plot_pars_b1_b3_b5_bb} (b1, b3, b5) and for the 7 long bursts in
Figs.~\ref{plot_pars_long_b2_b4_b7-b10_bb_log} (b2, b4, b7--b10)
and \ref{plot_pars_b6_bb_all_plus_fbolvst} (left panel; b6).
For the long bursts we used a logarithmic time scale, to emphasize
the start of the bursts, where changes in the parameters are most rapid.

\begin{figure}
 \resizebox{\hsize}{!}{\includegraphics[angle=-90, clip, bb=112 18 571 550]{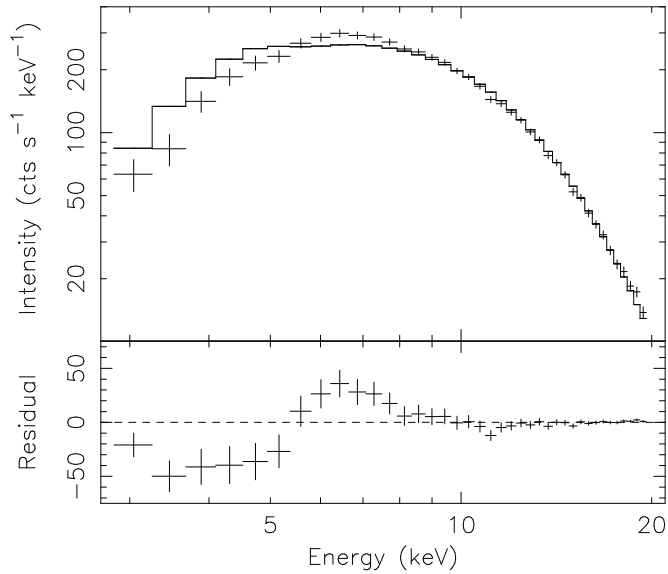}}
\caption{{\it Top panel}: 
Average observed net burst (i.e.\ total source minus persistent emission) 
spectrum during the flat top of burst b6,
i.e.\ from 27 to 187\,s after the start of the burst. A black-body fit it shown 
(kT$_{\rm bb}$=2.658$\pm$0.015\,keV, R$_{\rm bb,10}$=4.65$\pm$0.06\,km, $\chi^2_{\rm red}$=2.4 for 37 dof).
{\it Bottom panel}: Residuals after subtracting the best fit black-body model from the observed 
spectrum.}
\label{b6_flattop}
\end{figure}

The fits are, while much improved overall, still not optimal during the peaks of the long bursts b4, b6 and b10 
($\chi^2_{\rm red}$$=$2--3 with 44 [b4] or 37 [b6,b10] dof for the spectral fits to the 16\,s spectra). 
In Fig.~\ref{b6_flattop} we show the average net burst spectrum during the 
flat top part of burst b6 excluding the radius expansion and initial contraction phase (Sect.~3.2.4), 
i.e.\ 27--187\,s after the start of the burst. T$_{\rm bb}$ and R$_{\rm bb,10}$ do not 
change much during this interval. Clearly, deviations occur below $\simeq$10\,keV;
the intensity drops below $\simeq$5.5\,keV, while there is an excess between $\simeq$5.5--8\,keV.
These deviations are much larger than the calibration uncertainties.

In the short bursts T$_{\rm bb}$ decreases during the decay, indicating cooling
of the neutron star, as already noted from the hardness curves.
There are some slight variations in R$_{\rm bb,10}$
and T$_{\rm bb}$ during the first few seconds.
However, considering the behaviour of the net burst black-body flux, 
F$_{\rm bb,net}$, they do not show the correlations that would be expected for radius expansion events,
i.e.\ an increase in R$_{\rm bb,10}$ with a simultaneous drop in T$_{\rm bb}$ and a (nearly) constant 
F$_{\rm bb,net}$. 

The long bursts all show more or less the same behaviour (note again that
for burst b2 we have no short time scale spectral information, 
see Fig.~\ref{plot_pars_long_b2_b4_b7-b10_bb_log}). After a fast increase 
($\lesssim$0.5\,s) to maximum, F$_{\rm bb,net}$ remains constant 
(1.4--1.7$\times$10$^{-8}$\,erg\,s$^{-1}$\,cm$^{-2}$ for the various bursts) 
for 50--200\,s, and then decays exponentially. 
The duration of this flat-topped phase is similar to that seen in the light curve at high energies
(see middle panels of Figs.~\ref{plot_b2_b4_b6} and \ref{plot_b7-b10}).
During the first part T$_{\rm bb}$ increases, then it levels off, to 
decrease again when F$_{\rm bb,net}$ decreases. 

R$_{\rm bb,10}$ is $\simeq$5\,km during the decay of the burst, until
F$_{\rm bb,net}$ drops below typically 0.2$\times$10$^{-8}$\,erg\,s$^{-1}$\,cm$^{-2}$.
Then R$_{\rm bb,10}$ decreases again. 
Note that this latter behaviour resembles the systematic effects described by van Paradijs \&\ Lewin (1986)
near the end of the burst, when the persistent emission becomes important with respect to 
the net burst emission (Sect.~3.2.2.1). However, we have argued above that the burst black-body
emission is decoupled from the persistent black-body emission. It is unlikely that 
this behaviour is due to 
the fact that we assumed a constant persistent component during the bursts, while in fact
the persistent component may vary on the same time scale of the burst or faster
(see Sect.~3.1). If that was the case, one would expect that in some bursts R$_{\rm bb}$ would decrease, 
whereas in some others it would increase. 
On the contrary, we find that R$_{\rm bb}$ decreases in {\it all} bursts.
We will discuss this issue in Sect.~5.2.2.

\subsubsection{Radius expansion bursts}

The most dramatic changes in the spectral parameters can be seen during the first few seconds of the 
long bursts: while
there is a short dip in T$_{\rm bb}$ near the start of the burst, R$_{\rm bb,10}$
shows values which are up to a factor of $\simeq$6 larger than those found later on
(although this is not true for burst b10, see 
Fig.~\ref{plot_pars_long_b2_b4_b7-b10_bb_log}).
This is typical of radius expansion/contraction episodes, as described in Sect.~1.
Burst b6 shows this behaviour clearest, and may have had even two
of such episodes within one second (Fig.~\ref{plot_pars_b6_bb_all_plus_fbolvst}, left panel).
The expansion to maximum radius takes place in less than $\simeq$0.5\,s, while 
the initial contraction phases are longer: it takes  between 5 and 20\,s
from maximum expansion back to close to the neutron star surface.
The final contraction phase lasts between $\simeq$20 and $\simeq$180\,s
We conclude that bursts b4, and b6--b9 are radius expansion bursts.

In Fig.~\ref{plot_peak_pers}b we show the (average) 
net-burst peak fluxes, F$_{\rm bb,net\_peak}$,
determined in the same way as in the previous subsection for F$_{\rm bb,peak}$, 
but now for the net-burst emission. 
Most of the net burst peak fluxes are in the same range. For the radius 
expansion bursts this is expected, if the Eddington limit is reached.
Note that F$_{\rm bb,peak}$$\simeq$F$_{\rm bb,net\_peak}$ for bursts b6 and b7, as expected
if there is no black-body contribution to the persistent emission.
We conclude that F$_{\rm bb,net\_peak}$ in the radius expansion bursts
is a measure of the Eddington luminosity observed at Earth. 
We find that F$_{\rm bb,net\_peak}$ for the non-radius expansion bursts 
(except bursts b3) is more or less
similar to the maximum flux for the radius expansion bursts. 

Radius expansion events are most conveniently displayed in a flux-temperature diagram
(where temperature increases from right to left, analogous to HR-diagrams). 
Our best example, burst b6, is displayed this way in the
right-hand panel of Fig.~\ref{plot_pars_b6_bb_all_plus_fbolvst}. In this diagram
the radius expansion/contraction phase and the subsequent cooling of the neutron star 
are clearly distinguished by two separate tracks (see e.g.\ Lewin et al.\ 1993).
The data points distributed along the horizontal line in the upper part of the diagram 
(i.e.\ nearly constant F$_{\rm bb,net}$ of 1.2--1.3$\times$10$^{-8}$\,erg\,s$^{-1}$\,cm$^{-2}$)
represent the expansion/contraction phase. The data points distributed along the
diagonal line from the upper left to the lower right part of the diagram are from the 
cooling phase of the burst. 
We note that the point in between the two tracks is from the very first rise phase.
Comparing both panels of Fig.~\ref{plot_pars_b6_bb_all_plus_fbolvst}
it is apparent that the burst spends a long time near the vertex of both tracks, about 150\,s,
where F$_{\rm bb,net}$$\simeq$1.24$\times$10$^{-8}$\,erg\,s$^{-1}$\,cm$^{-2}$ and 
kT$_{\rm bb}$$\simeq$2.65\,keV. 
The first part (F$_{\rm bb,net}$$\gtrsim$0.4$\times$10$^{-8}$\,erg\,s$^{-1}$\,cm$^{-2}$)
of the cooling track is well described
($\chi^2_{\rm red}$=0.6 for 9 dof) by a straight line 
with a slope of 4.14$\pm$0.15 (dotted line in Fig.~\ref{plot_pars_b6_bb_all_plus_fbolvst}), i.e.\
F$_{\rm bb,net}$ is consistent with being proportional to
T$_{\rm bb}^4$, as expected if the neutron star photosphere
radiates as a black-body with constant effective area.
Below F$_{\rm bb,net}$$\simeq$0.4$\times$10$^{-8}$\,erg\,s$^{-1}$\,cm$^{-2}$ the data
start to deviate from this relation.
The flux-temperature diagrams for the other long bursts are more or less
consistent with the behaviour of burst b6, although the exact locations of 
the vertices between the radius expansion/contraction phase and the
cooling phase differ slightly (kT$_{\rm bb}$$\simeq$2.3--2.6\,keV, 
F$_{\rm bb,net}$$\simeq$1.2--1.35$\times$10$^{-8}$\,erg\,s$^{-1}$\,cm$^{-2}$). 
The early parts of the cooling phases for the other long bursts are in most cases also consistent
with black-body cooling, although the flux level at which they start to deviate 
from F$_{\rm bb,net}$$\propto$T$_{\rm bb}^4$ varies from burst to burst 
(e.g.\ $\simeq$1$\times$10$^{-8}$\,erg\,s$^{-1}$\,cm$^{-2}$ for burst b4). The slope
becomes somewhat steeper below these fluxes for all bursts. 

\subsubsection{The radius expansion/contraction track}

\begin{figure}
 \resizebox{\hsize}{!}{\includegraphics[angle=-90, clip, bb=50 80 566 588]{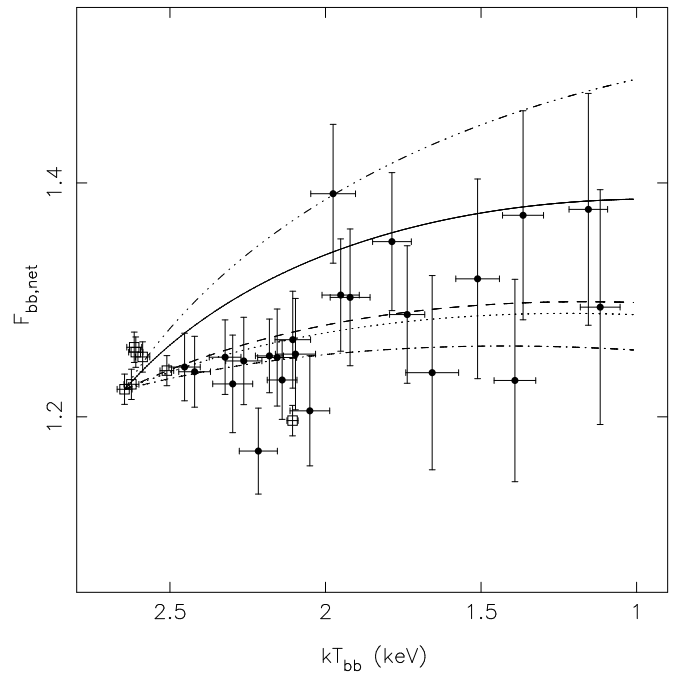}}
\caption{Same as the right panel Fig.~\ref{plot_pars_b6_bb_all_plus_fbolvst}, but only the radius 
expansion/contraction phase is shown (note that the highest flux point lies outside the plot boundaries). 
The five lines correspond to expected tracks in the same diagram
for different sets of parameters, see text.}
\label{plot_b6_fbolvst_parameters}
\end{figure}

As we have shown above, the observed net burst flux is consistent with being constant during the radius 
expansion/contraction phases. For instance, for 
burst b6 a constant fit yields 
F$_{\rm bb,net}$$\simeq$1.24$\times 10^{-8}$\,erg\,s$^{-1}$\,cm$^{-2}$ 
with a reduced $\chi^2$ of 1.4 for 29 dof, excluding the highest flux point. 
However, for a constant composition and constant 
(an)isotropy of the radiation, one would expect to see a slight increase in the observed
bolometric burst flux with increasing photosphere radius, since the gravitational redshift 
decreases (see e.g.\ Lewin et al.\ 1993).
In Fig.~\ref{plot_b6_fbolvst_parameters}, we zoom in on the observed radius expansion/contraction phase
of burst b6 (note that for F$_{\rm bb,net}$ and T$_{\rm bb}$ we now use a linear scale). 
Overdrawn is an example of the expected relation between bolometric burst flux and 
the photospheric radius (continuous line), for `standard' values of 
the mass of the neutron, $M_{\rm ns}$=1.4\,M$_{\sun}$, the hydrogen fraction
(by mass), $X$=0.73, the spectral hardening factor, 
T$_{\rm bb}$/T$_{\rm eff,\infty}$=1.7, and high-temperature electron
scattering opacity, see Appendix~B for details.
The curve was normalized so as to give the observed values of
F$_{\rm bb,net}$ and T$_{\rm bb}$ at touch down (the leftmost point of the 
radius expansion/contraction track). In practice this is analogous to solving for the distance and 
the photospheric radius R at touch down (see Appendix~B). 
We can see that the curve does not follow most of the data points.
Taking the low-temperature electron scattering opacity 
(but keeping the other parameters at their
`standard' values, see above) makes the disagreement even larger 
(dash-dot-dot-dotted line in Fig.~\ref{plot_b6_fbolvst_parameters}). 
There are three ways to flatten the radius expansion/contraction track, or
equivalently, to reduce the gravitational redshift corrections.
The most obvious one is to lower the mass of the neutron star; the expected track (dash-dotted
line) for $M_{\rm ns}$=0.5\,M$_{\sun}$ is given. However, such a low-mass neutron star 
is in not in line with the observed and expected masses for neutron stars
(see e.g.\ Thorsett \&\ Chakrabarty 1999).
A second option is to lower the hydrogen content of the burning material
(shown by the dashed line for $X$=0). However, the long contraction phase is typical for
unstable mixed H/He-burning (see Sect.~5.4) 
and not expected for unstable pure He-burning. Moreover, the distance derived
at touch down is then rather high (d=15.2\,kpc for $X$=0, see Appendix B). 
Finally, increasing the hardening factor to T$_{\rm bb}$/T$_{\rm eff,\infty}$=2
also flattens the expected track (dotted line in Fig.~\ref{plot_b6_fbolvst_parameters}).
This may be the most realistic option, since 
spectral hardening values of $\sim$2 are inferred for burst luminosities near the Eddington limit 
(Babul \&\ Paczy\'nski 1987; but see Titarchuk 1994, where T$_{\rm bb}$/T$_{\rm eff,\infty}$$<$1 
at very large photospheric radii). 
We note that the effects of an expanded boundary layer, as discussed in Sect.~5.2.1, may also
be of importance here. However, at present it is unclear how to take this into account.

\subsubsection{Burst parameters}

From X-ray spectral fits
we can determine, for each burst, the maximum net bolometric black-body flux, 
F$_{\rm bb,max}$, and the total burst fluence
(i.e.\ the integrated net burst flux), E$_{\rm b}$. These can be used to derive the burst parameters
$\gamma$=F$_{\rm pers}$/F$_{\rm bb,max}$, where F$_{\rm pers}$ is the persistent flux
flux between 2 and 20\,keV (Table~3), and the average burst duration, $\tau$=E$_{\rm b}$/F$_{\rm bb,max}$.
Note that F$_{\rm bb,max}$ may be slightly underestimated due to the finite width of the time bins.
The magnitude of this effect depends on how fast the flux varies with time.
E$_{\rm b}$ has been determined by adding up the observed net burst fluxes, F$_{\rm bb,net}$, per time bin
from the beginning to the estimated end of the burst.
Since F$_{\rm bb,net}$ decays exponentially it only vanishes finishes at infinite times.
To compensate for this we fit the decay with an exponential and determine
the `rest' fluence by integrating from the estimated end of the burst up to infinity.
The `rest' fluence is large only for burst b2, which was interrupted during its decay
(see Fig.~\ref{plot_b2_b4_b6}).
We also determined the burst parameter $\alpha$, which is the ratio of the average persistent flux to the 
time-averaged flux emitted in the bursts, $\alpha$=F$_{\rm pers}$/(E$_{\rm b}$/$\Delta$t), 
where $\Delta$t is the time since the previous burst.
For $\alpha$ we can only give lower limits, 
since the source is not observed during South Atlantic Anomaly passages and earth occultations. 
For burst b5 we have, however, also assumed that between bursts b4 and b5 no other bursts
occurred ($\Delta$t=5.77\,hr). Note that when we derive $\alpha$ we assume that between
bursts the persistent luminosity is constant. Since GX\,17+2 is a highly variable source
(see e.g.\ Fig.~\ref{lc_b10}), this is not strictly valid.
The burst parameters can be found in Table~4.

\begin{table*}
\caption{Burst parameters}
\begin{tabular}{lccccl}
\hline
\multicolumn{1}{c}{burst} & \multicolumn{1}{c}{F$_{\rm bb,max}$$^a$} &
\multicolumn{1}{c}{E$_{\rm b}$$^b$} &  \multicolumn{1}{c}{$\tau$ (s)} &
\multicolumn{1}{c}{$\gamma$} & \multicolumn{1}{c}{$\alpha$} \\
\hline
b1     & 1.59$\pm$0.06 & 5.38$\pm$0.07 & 3.39$\pm$0.14 & 1.3$\pm$0.2 & $\gtrsim$700 \\
b2$^c$ & 1.33$\pm$0.01 & 356.4$\pm$0.7 & 267$\pm$2     & 1.7$\pm$0.4 & $\gtrsim$15 \\
b3     & 0.79$\pm$0.05 & 5.54$\pm$0.09 & 7.0$\pm$0.5   & 2.4$\pm$0.4 & $\gtrsim$250 \\
b4     & 1.59$\pm$0.06 & 430.2$\pm$1.1 & 271$\pm$11    & 1.2$\pm$0.2 & $\gtrsim$10\\
b5     & 1.46$\pm$0.07 & 5.26$\pm$0.07 & 3.60$\pm$0.17 & 1.6$\pm$0.2 & $\gtrsim$100 \\
b5     & \multicolumn{4}{c}{~} & \multicolumn{1}{l}{9000$\pm$1300$^d$} \\
b6     & 1.68$\pm$0.09 & 664.1$\pm$1.9 & 396$\pm$22    & 1.5$\pm$0.1 & $\gtrsim$1 \\
b7     & 1.36$\pm$0.08 & 143.3$\pm$1.0 & 106$\pm$6     & 1.8$\pm$0.1 & $\gtrsim$5 \\
b8     & 1.53$\pm$0.08 & 176.4$\pm$1.1 & 115$\pm$6     & 1.6$\pm$0.3 & $\gtrsim$25 \\
b9     & 1.50$\pm$0.08 & 137.6$\pm$1.0 & 92$\pm$5      & 1.6$\pm$0.3 & $\gtrsim$25 \\
b10    & 1.65$\pm$0.08 & 169.0$\pm$0.9 & 102$\pm$5     & 1.2$\pm$0.2 & $\gtrsim$15 \\
\hline
\multicolumn{6}{l}{\footnotesize $^a$\,Bolometric peak net-burst black-body flux in 10$^{-8}$\,erg\,s$^{-1}$\,cm$^{-2}$.} \\
\multicolumn{6}{l}{\footnotesize $^b$\,Burst fluence in 10$^{-8}$\,erg\,cm$^{-2}$.} \\
\multicolumn{6}{l}{\footnotesize $^c$\,Using the 16\,s spectral fit results.} \\
\multicolumn{6}{l}{\footnotesize $^d$\,Assuming that between burst b4 and b5 no other bursts occurred.}\\
\label{burst_parameters}
\end{tabular}
\end{table*}

\subsection{Burst position in the Z}

\begin{figure*}
 \resizebox{6.43cm}{!}{\includegraphics[angle=0, clip, bb=20 154 498 621]{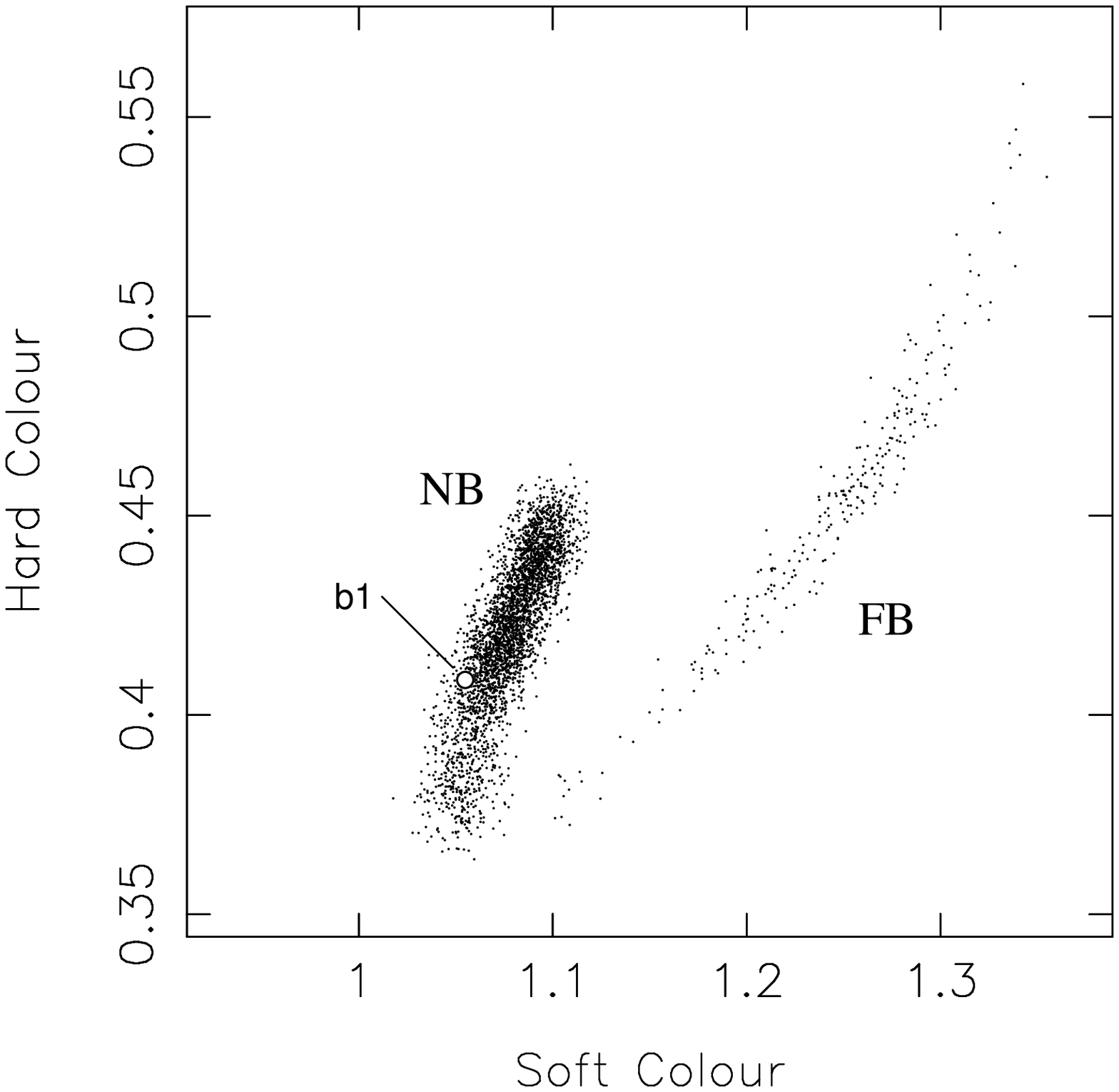}}
 \resizebox{5.785cm}{!}{\includegraphics[angle=0, clip, bb=68 154 498 621]{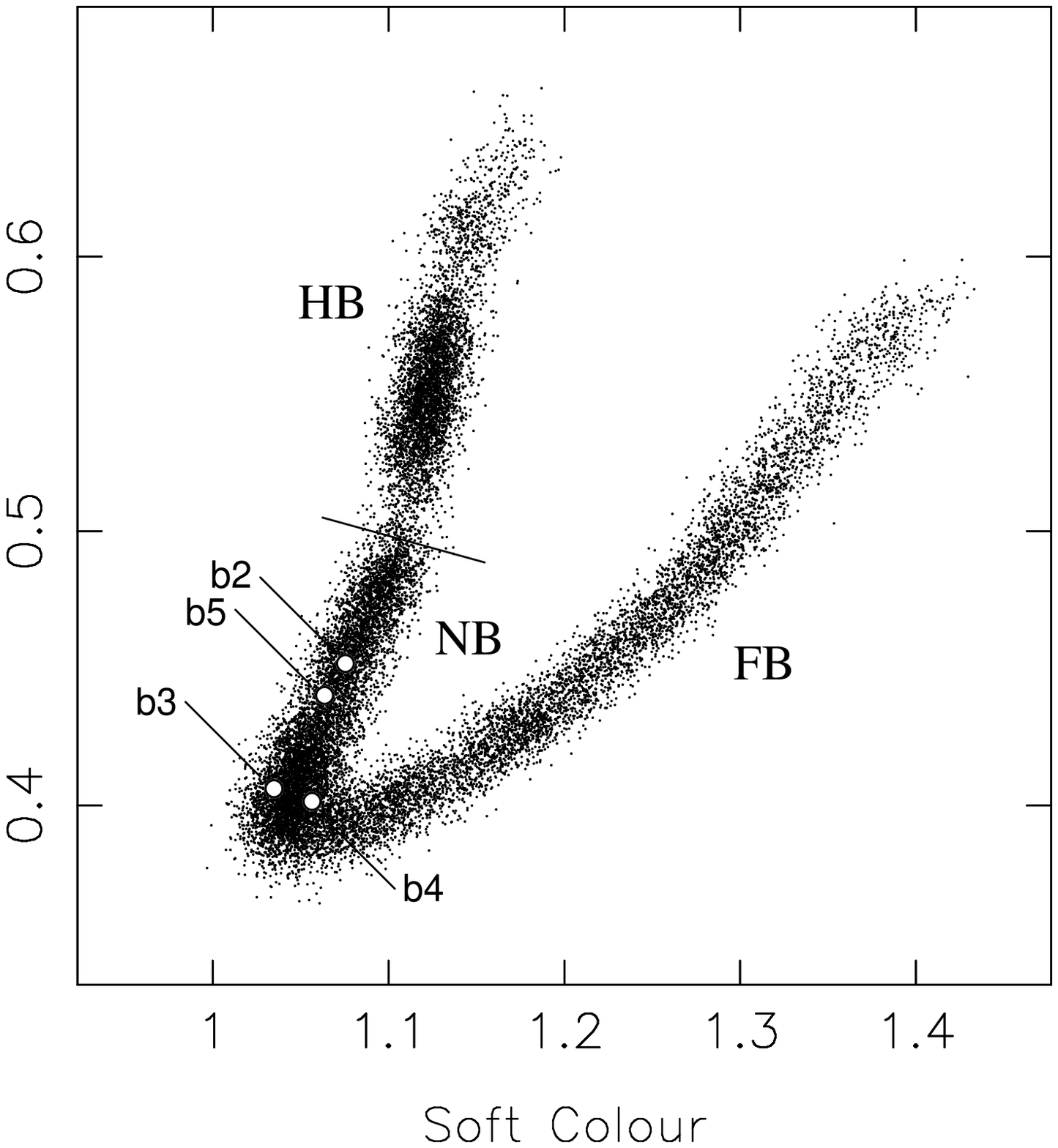}}
 \resizebox{5.785cm}{!}{\includegraphics[angle=0, clip, bb=68 154 498 621]{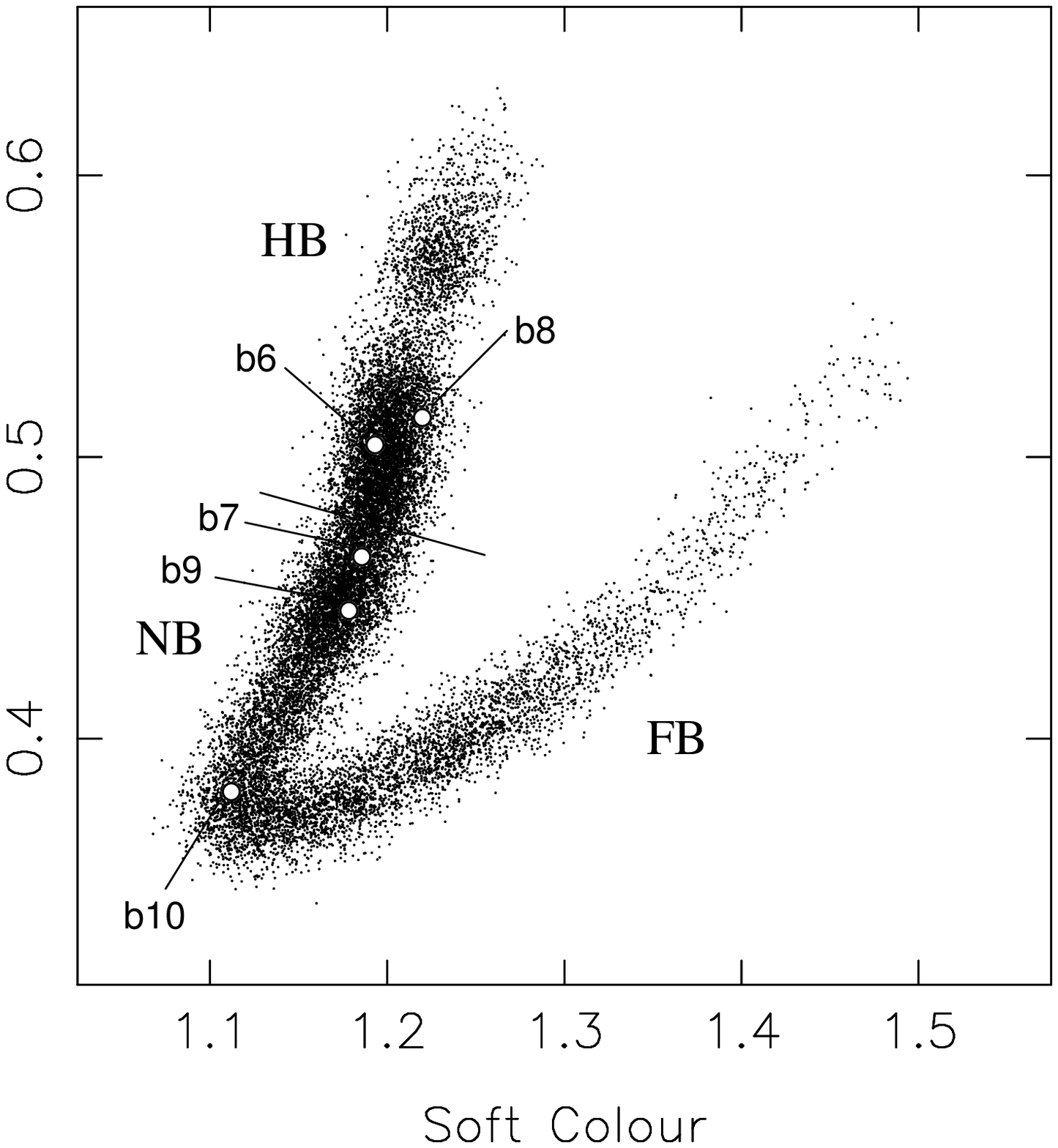}}
\caption{Colour-colour diagrams of the RXTE/PCA observations of GX\,17+2 (after Homan et al.\ 2001).
The left, middle and right panels refer to the observations done during RXTE gain epochs 1, 3 and 4,
respectively. 
The soft colour is defined as the count rate ratio in the 4.8--7.2\,keV to the 3.0--4.8\,keV 
energy bands for the epoch 1 data, in the 4.8--7.3\,keV to the 3.0--4.8\,keV 
energy bands for the epoch 3 data, and in the 4.6--7.1\,keV to the 3.0--4.6\,keV energy bands 
for the epoch 4 data. The hard colour is defined as the count rate ratio in the 10.6--19.7\,keV 
to the 7.2--10.6\,keV energy bands for the epoch 1 data, in the 10.5--19.6\,keV to the 7.3--10.5\,keV 
energy bands for the epoch 3 data, and in the 10.5--19.5\,keV to the 7.1--10.5\,keV energy bands 
for the epoch 4 data. Each point represents an average of 16\,s.
We have indicated the position of the source during 64\,s intervals just before a burst
(indicated by b1, b2, ..., b10, in chronological order), and the three limbs of the Z (HB, NB, FB).
Since the HB is not clearly distinguishable from the NB in this figure, we have indicated the approximate
HB/NB vertex by a line.}
\label{cd}
\end{figure*}

The colour-colour diagram (CD) of the data from the three different RXTE gain epochs is 
shown in Fig.~\ref{cd},
together with the source positions just before the burst (see also Table~2). 
To determine the position of the source in the CD at the time of the burst, we calculated the
soft and hard colour values from 64\,s intervals just before a burst.
The three short bursts all occurred when the source was in the lower part of the NB. 
Among the long bursts, two occurred in the lower part 
of the HB, four in the NB, and one close to the NB/FB vertex. 
During the observations in 1996--2000 GX\,17+2 spent about 28\%\ of its time 
in the HB, 44\%\ of its time in the NB, and 28\%\ in the FB.
It thus seems that the NB is overpopulated, whereas the FB is underpopulated 
with bursts (i.e.\ none occurred).
However, by assuming that bursts have an equal chance to occur at each instant independent of branch,
this result is statistically not significant (4\%\ probability seeing 7 out of 10 bursts in the NB, and 
7\%\ of seeing none out of 10 bursts in the FB).
Our conclusion does not change much if we include the EXOSAT/ME results reported by 
Kuulkers et al.\ (1997);
for 260\,ksec of observing time, GX\,17+2 spent 11\%\ of the time in the HB, 66\%\ in the NB and 
23\%\ in the FB, with 3 bursts occurring in the NB and 1 in the (lower part of the) FB.
This leads to total 
probabilities of 7\%\ of seeing 10 out of 14 bursts in the NB, and 6\%\ of seeing only 1 out of 14 bursts
in the FB.

\section{Search for burst oscillations}

Following the method described by Leahy et al.\ (1983; see also Vaughan et al.\ 1994) 
we defined a 99\%\ confidence level above which
powers are regarded to be due to a real signal. This level depends on
the number of trials, which basically is the number of independent
frequency bins that are examined. In our search this number is
$\simeq$2$\times$10$^7$ which leads to a trigger level of 42.8 in the Leahy et al.\ (1983)
normalization.
This level was never reached, indicating that no coherent oscillations were
detected in any of the bursts. We inspected the frequencies of the highest observed 
powers. These powers
occurred at apparently random frequencies, consistent with the idea that they are due
to random fluctuations.

Since no powers were found above the trigger level, only 
upper limits to the strength of possible burst oscillations can be given
(see van der Klis 1989; Vaughan et al.\ 1994). 
We used the (maximum) observed powers to determine a 99\%\ confidence 
upper limit to a true signal power.
These upper limit powers were then converted 
into the full amplitude of an assumed sinusoidal 
signal\footnote{If a sinusoidal signal is given by $y=a\sin x$, the full amplitude is defined
as $A=2a$.}, which is given by:
\begin{eqnarray}
A=1.61 \sqrt{\frac{P_{\rm ul}}{N}} \left[ {\rm sinc} \left( \frac{\pi}{2} \frac{\nu}{\nu_{\rm Nyq}} \right) \right]^{-1}, \nonumber
\end{eqnarray}
where $P_{\rm ul}$ is the (upper limit) signal power (Vaughan et al.\ 1994; Groth 1975), $N$ the
number of photons used in the power spectrum, $\nu$ the frequency at which the
power was found, and $\nu_{\rm Nyq}$ the Nyquist frequency. Table~\ref{upper_tab} 
gives the upper limits on $A$ for all bursts --- each
burst was divided into four parts: 5\,s before the rise, the rise, the
peak, and decay (= end of peak + 2$\times$t$_{\rm exp}$). Note that $A$ is a factor
$\sqrt{2}$ larger than the rms amplitude that is often quoted.

The values of the upper limits are not very constraining. Only those
obtained from the 2\,s power spectra in the 2--60\,keV band are well below
the strongest values observed in some of the atoll sources
(see e.g.\ Strohmayer 1998, 2001).

\begin{table*}
\caption{Upper limits on burst oscillations}
\begin{tabular}{cccccccccccc}
\hline
burst & $\phi$$^a$ & \multicolumn{4}{c}{upper limits on fractional amplitude$^b$} & 
burst & $\phi$$^a$ & \multicolumn{4}{c}{upper limits on fractional amplitude$^b$} \\
      &            & 0.25\,s      & 2\,s         & 0.25\,s      & 2\,s &
      &            & 0.25\,s      & 2\,s         & 0.25\,s      & 2\,s \\
      &            & 2--60\,keV & 2--60\,keV & 8--20\,keV & 8--20\,keV &
      &            & 2--60\,keV & 2--60\,keV & 8--20\,keV & 8--20\,keV \\
\hline
b1$^c$&  b    & 0.28 & 0.12  & $>$1.0  & 0.67 & b6   &  b     & 0.41 & 0.14  & 0.70    & 0.28 \\
     &  r     & 0.23 & 0.07  & $>$1.0  & 0.34 &      &  r     & 0.28 & 0.09  & 0.49    & 0.24 \\
     &  p     & 0.25 & 0.09  & $>$1.0  & 0.45 &      &  p     & 0.40 & 0.15  & 0.63    & 0.24 \\
     &  d     & 0.27 & 0.11  & $>$1.0  & 0.60 &      &  d     & 0.45 & 0.16  & 0.89    & 0.30 \\
b2   &  b     & 0.34 & 0.10  & 0.66    & 0.26 & b7   &  b     & 0.47 & 0.16  & 0.78    & 0.26 \\
     &  r     & 0.19 & 0.07  & 0.51    & 0.20 &      &  r     & 0.30 & 0.08  & 0.58    & 0.15 \\
     &  p     & 0.29 & 0.10  & 0.55    & 0.20 &      &  p     & 0.36 & 0.13  & 0.63    & 0.23 \\
     &  d     & 0.33 & 0.11  & 0.68    & 0.26 &      &  d     & 0.46 & 0.16  & 0.80    & 0.34 \\
b3   &  b     & 0.36 & 0.13  & 0.77    & 0.25 & b8   &  b     & 0.41 & 0.14  & 0.74    & 0.26 \\
     &  r     & 0.21 & 0.08  & 0.51    & 0.17 &      &  r     & 0.33 & 0.10  & 0.56    & 0.20 \\
     &  p     & 0.32 & 0.12  & 0.64    & 0.24 &      &  p     & 0.37 & 0.14  & 0.59    & 0.24 \\
     &  d     & 0.32 & 0.12  & 0.71    & 0.25 &      &  d     & 0.43 & 0.16  & 0.72    & 0.27 \\
b4   &  b     & 0.37 & 0.14  & 0.76    & 0.27 & b9   &  b     & 0.41 & 0.13  & 0.77    & 0.30 \\
     &  r     & 0.30 & 0.09  & 0.62    & 0.23 &      &  r     & 0.25 & 0.09  & 0.56    & 0.22 \\
     &  p     & 0.29 & 0.11  & 0.55    & 0.20 &      &  p     & 0.38 & 0.14  & 0.63    & 0.22 \\
     &  d     & 0.38 & 0.13  & 0.75    & 0.27 &      &  d     & 0.45 & 0.17  & 0.78    & 0.28 \\
b5   &  b     & 0.33 & 0.11  & 0.63    & 0.22 & b10  &  b     & 0.45 & 0.19  & 0.67    & 0.34 \\
     &  r     & 0.24 & 0.09  & 0.42    & 0.17 &      &  r     & 0.34 & 0.09  & 0.48    & 0.23 \\
     &  p     & 0.27 & 0.10  & 0.39    & 0.17 &      &  p     & 0.38 & 0.14  & 0.51    & 0.26 \\
     &  d     & 0.28 & 0.10  & 0.57    & 0.22 &      &  d     & 0.48 & 0.18  & 0.68    & 0.34 \\
\hline	 
\multicolumn{12}{l}{\footnotesize $^a$\,Phase of burst profile: b=before burst, r=rise, p=peak, d=decay.} \\
\multicolumn{12}{l}{\footnotesize $^b$\,99\%\ confidence upper limits in the 50--2000 Hz frequency range.}\\
\multicolumn{12}{l}{\footnotesize $^c$\,99\%\ confidence upper limits in the 2--60\,keV energy band are 
between 50--250\,Hz. The high-energy band used is 13.5--20\,keV.}\\
\end{tabular}	 
\label{upper_tab}
\end{table*}

\section{Discussion}

\subsection{X-ray burst properties} 

Of the ten X-ray bursts from GX\,17+2 seen so far with the RXTE/PCA, three are rather short ($\simeq$10\,s),
whereas the other seven have durations of $\simeq$6--25\,min. 
They all show evidence for cooling of the neutron star photosphere during the decay,
confirming previous results.
The burst durations found are in the range seen previously (Kahn \&\ Grindlay 1984; 
Tawara et al.\ 1984c; Sztajno et al.\ 1986; Kuulkers et al.\ 1997). Using the time-resolved X-ray 
spectral fits (see Sect.~5.2) we determined the 
burst fluence, $\tau$ (a characterisation of the burst decay time), $\gamma$ 
(the net brightness of the burst with respect to the persistent level) and $\alpha$ (the ratio
between the total net burst flux and total persistent flux in between bursts).
Since not all these parameters were given for GX\,17+2 bursts observed previously, we 
(re-)\-determined
these values from the information given in the various papers. This is explained in 
Appendix~A and the results are given in Table~A.1. 
Again, our values for the burst parameters
for both the short and long bursts are comparable to those for the bursts observed previously. 
Since observations with RXTE are interrupted by either SAA passages or Earth occultations,
and the fact that the length of each `continuous' observation was limited ($\sim$hours)
we could not determine the time interval between bursts. 
Our lower limits on $\alpha$ are, therefore, rather low. However, from the bursts observed previously, it is 
found that $\alpha$ always exceeds $\sim$1000. 
This means that most of the accreted material is being burned continuously. 
One of our short bursts occurred about 5.8\,hr after a long burst.
So far this is the shortest recurrence time encountered for a burst in GX\,17+2.
The fluences of the bursts from GX\,17+2 are either $\simeq$5.5$\times$10$^{-8}$\,erg\,cm$^{-2}$
(short bursts) or 140--670$\times$10$^{-8}$\,erg\,cm$^{-2}$ (long bursts). At a
distance of 10\,kpc (but see Sect.~5.3) these correspond to total energies of 
6.5$\times$10$^{38}$\,erg and
2--8$\times$10$^{40}$\,erg, respectively, in the range
seen for type~I bursts (see e.g.\ Lewin et al.\ 1993). 

One of the short bursts was rather faint, with a peak intensity which was a factor of $\simeq$1.25 lower than for
the other bursts, and displayed two peaks in the light and bolometric flux curves. 
This burst resembles the (relatively) weak double-peaked bursts seen from 4U\,1636$-$53
(Sztajno et al.\ 1985; Fujimoto et al.\ 1988). Such bursts are clearly not radius expansion/contraction
events, but are instead thought to be due to the mixing of fresh material into the 
unstably burning layer; such mixing may be caused by shear instabilities in the outer envelope of 
the neutron star (Fujimoto et al.\ 1988). 

We show for the first time that almost all of the long bursts of GX\,17+2 
have episodes of radius expansion and 
contraction of the neutron star photosphere.
In all cases the expansion and initial contraction occur during the first 
few seconds of the bursts, but 
the final contraction phase duration varies between $\simeq$20\,s 
and $\simeq$180\,s. During the expansion/contraction phase the 
net burst flux remains constant, presumably at the Eddington value.
Long contraction phases might be caused by long phases of unstable hydrogen burning,
whose time scale is limited by $\beta$-decays, up to $\sim$100\,s after the burst has started
(see e.g.\ Bildsten 1998, 2000, and references therein). 
The maximum apparent radius at 10\,kpc, R$_{\rm bb,10}$, 
reached during the expansion episodes is $\simeq$30\,km.
Corrected for gravitational redshift effects and spectral hardening (see Appendix B), 
the maximum photospheric radius R (as measured by a local observer) is $\simeq$90--125\,km. 
We do not see evidence for a change in the cut-off power-law component above $\sim$10\,keV
when the photosphere expands up to these radii. 
This suggests that either the environment in between the neutron star surface
and expanded photosphere at $\simeq$90--125\,km is involved in producing the high-energy radiation
but was not affected, or the inner disk and corona which presumably produce the high-energy 
radiation do not extend down to the neutron star. 
A possibility is that the magnetospheric radius extends out to at least
these radii, disrupting the disk and corona near that point. Note that the
radius range of $\simeq$90--125\,km overlaps with the range of magnetospheric radii
expected for Z sources 
($\sim$10--100\,km, see Ghosh \&\ Lamb 1992). However, at the high mass accretion rates
inferred for GX\,17+2 the disk is expected to extend down to close to the neutron star 
(see Popham \&\ Sunyaev 2001). For a more detailed discussion of the different
emission regions we refer to the next subsection.

Long ($>$minutes) X-ray bursts have been seen in many other sources 
(see, in order of decreasing duration, 
Hoffman et al.\ 1978b, Lewin et al.\ 1984, $\simeq$25\,min: 4U\,1708$-$23;
Swank et al.\ 1977, $\gtrsim$10\,min: 4U\,1724$-$307; 
Tawara et al.\ 1984a,b, $\gtrsim$4.5\,min: 3A\,1715$-$321;
Kaptein et al.\ 2000, $\gtrsim$3.3\,min: 1RXS\,J171824.2$-$402934;
van Paradijs et al.\ 1990, $\gtrsim$2.7\,min: 4U\,2129+11).
The shapes of the 
light curves of these bursts at different energies are very similar to the long bursts of GX\,17+2,
i.e.\ fast (several seconds) rise and peaked emission at low energies and relatively slower 
rise at high energies. They also
have rather long episodes of contraction, i.e.\ long episodes of Eddington fluxes, up to
$\simeq$300\,s (see e.g.\ Lewin et al.\ 1984). Moreover, 
the curves for T$_{\rm bb}$ and R$_{\rm bb}$ as a function of time
show similar shapes. 

The difference, however, between the long bursts of GX\,17+2 and those in the other sources is
that `precursors' are present in the light curves before almost all of the long 
bursts from other sources. Note that these are not
real precursors, but indicate very large radius expansion (R$_{\rm bb}$$>$100\,km).
No emission is then seen because of the very low effective temperatures reached; the
peak of the emission is shifted to UV wavelengths. Thus the X-ray burst light curve is
interrupted by a gap of no X-ray emission leading to an apparent precursor 
(see Lewin et al.\ 1984). The rise to maximum radius in those sources is on the order
of several seconds. In GX\,17+2 the maximum value of R$_{\rm bb}$ is a few times the 
neutron star radius and the expansion lasts typically only a second.
The most important difference, however, is that 
the persistent emission just before the long bursts in other sources 
is $\sim$1\%\ of the Eddington value, i.e.\ a factor of $\sim$100 lower than in GX\,17+2
(see also Sect.~5.4).

We note that even longer X-ray bursts do exist, which last for several hours 
(Cornelisse et al.\ 2000, 2001; Wijnands 2001; Strohmayer \&\ Brown 2001; 
Kuulkers 2001; Kuulkers et al.\ 2001). Their recurrence 
time is much longer than normal type~I bursts ($\gtrsim$7.5~days, Cornelisse et al. 2001;
$\lesssim$4.8\,yrs, Wijnands 2001), and they seem to occur at persistent 
luminosites which are a factor of $\sim$0.1--0.2 of the Eddington limit. 

\subsection{Black-body spectral fits}

\subsubsection{Decoupled persistent and burst emission}

It has previously been pointed out (van Paradijs \&\ Lewin 1986; see also Sztajno et al.\ 1986) 
that, if the persistent emission before the burst
contains a spectral component which originates from the same regions as the burst emission,
this should be taken into account in 
the spectral analysis of X-ray bursts. 
This is especially the case when the neutron
star is accreting matter at high rates such as thought to be the case in Z sources, 
making the neutron star hot and radiate soft (black-body like) emission. 
`Standard' X-ray burst spectral analyses (modeling the net-burst emission by black-body radiation)
then results in systematic errors in the spectral parameters, especially during the end of the
burst when the net-burst luminosity is small. It was therefore suggested to model the total
burst emission using a two-component approach, i.e.\ with a soft component (usually a
black body) and a hard component (fixed at the parameters found for the persistent emission).
The former component is thought to contain all emission from the neutron star photosphere, while
the latter component is assumed to be from the accretion disk and not affected by the prompt
burst. However, we find that using this two-component spectral fitting approach does 
{\it not} give satisfactory results. 
Instead, the `standard' burst analysis, which assumes the persistent emission
continues unchanged, yield better fits. 

This suggests that the persistent emission is unaffected during the burst and 
that the persistent black-body component (which is not always present)
does not arise from the same region as the burst emission, contrary to what was previously assumed.
We find that during the initial exponential decay phase of the bursts the neutron star photosphere
shows pure black-body cooling. During that stage the 
apparent black-body radius stays constant at 5--6\,km (at 10\,kpc).
Comparing these radii with the apparent radii of the black-body component of 
the persistent emission (if present) indicates that the persistent black-body component originates 
from a larger area than that of the burst emission (see Fig.~\ref{plot_pars_b4log_b1_b4_b6}). 
The inner part of the accretion disk or an expanded boundary layer are
the most obvious sites for this persistent soft component.
A recent study of X-ray spectra from LMXBs with luminosities spanning several
orders of magnitude has shown that the persistent black-body emission likely does not arise in
(part of) the accretion disk (Church \&\ Ba\l uci\'nska-Church 2001).
Instead
at very high accretion rates (near Eddington values)
the accretion disk boundary layer around the neutron star expands radially, with
radial extents larger than one stellar radius (Popham \&\ Sunyaev 2001). 
Moreover, at these high rates the expected spectrum of
the boundary layer strongly resembles a black-body.
The fact that we see black-body emission in the persistent spectra with apparent 
radii which are larger (factor $\sim$2--3) 
than the neutron star (see also Church \&\ Ba\l uci\'nska-Church 2001)
is in line with this. Still, one would expect that such a boundary layer would be affected 
by the bursts (see Popham \&\ Sunyaev 2001), of which we do
not see clear evidence from our spectral fits.
We note, however, that Homan et al.\ (2001), who studied the properties of the NB and FB
quasi-periodic oscillations (NB/FB QPO or NBO/FBO) 
during bursts b4 and b10, found that during the 
bursts the absolute amplitude of these QPOs decreased 
significantly. This suggests that the inner accretion flow {\it is} affected by the 
increase in the radiation from the neutron star, contrary to what our spectral 
results indicate. Apparently the QPO mechanism of the NBO/FBO is much more 
sensitive to the radiation field than the bulk flow of matter itself. This is in 
accordance with models for NBO/FBO (Fortner et al. 1989), which require a 
delicate balance between the radiation field and the accretion flow, that might 
easily be disrupted during X-ray bursts.

\subsubsection{Deviations from black-body emission}

During the radius expansion/contraction phase black-body emission
does not provide a good description of our observed net-burst spectra, leaving a slight excess in
emission between $\simeq$5.5--10\,keV and lack of emission between $\simeq$3--5.5\,keV
(Fig.~11). Such systematic deviations
have been observed before, especially during radius expansion/contraction phases. 
For example, spectra obtained with the Large Area Counter onboard {\em Ginga} 
during the long burst of 4U\,2129+11 (van Paradijs et al.\ 1990) show
remarkably similar deviations at the same energies as we see in GX\,17+2.
The persistent mass accretion in 4U\,2129+11 is inferred to be a factor of
$\sim$100 lower than for GX\,17+2. This suggests that the 
radiation properties of the photospheres are the same when they are radiating near the 
Eddington luminosity, which is independent of the mass accretion rate.
As noted by van Paradijs et al.\ (1990), the `bumpiness' of the residuals
point to the presence of relatively narrow-band spectral features, rather 
than the residuals being due to broad-band spectral hardening at Eddington luminosities
(see also Strohmayer \&\ Brown 2001).

From our `standard' spectral analysis, we find that
the apparent black-body radii decrease near the end of the decay of the bursts. This was also inferred from
the EXOSAT/ME observations of GX\,17+2 (Sztajno et al.\ 1986), and has been seen 
in other sources as well (e.g.\ Chevalier \&\ Ilovaisky 1990).
If this were due to a real decrease
in the burst emitting area one might expect to see burst oscillations due to the
spinning neutron star. We do not see evidence for this.
If the expanded boundary layer reacts to its changing
environment on the same time scale as the burst itself then variations in
this layer's structure could also play a role here. 

This observed behaviour resembles the systematic effects
described by van Paradijs \&\ Lewin (1986), which occurs near the end of the bursts in the
presence of a persistent hot neutron star component. However, we have argued that the
persistent black body component does not contribute to the burst emission, so this is not a viable 
explanation. Sztajno et al.\ (1986) argued that this effect also can not be due to
spectral hardening, since that would lead to an apparent radius {\it increase} with 
decreasing T$_{\rm bb}$. Their argument was based on the assumption that the spectral
hardening decreases with decreasing temperatures (and therefore decreasing burst luminosity; see
also Sztajno et al.\ 1985). However, later model calculations have shown that in fact
spectral hardening {\it increases} again whenever
the burst luminosity drops below a certain value (L$_{\rm bb}$$\lesssim$0.2L$_{\rm Edd}$, 
e.g.\ London et al.\ 1986, Ebisuzaki 1987). For an assumed constant size of the emission region 
during the cooling stage (presumably the whole neutron star surface), the apparent radii will
then become smaller as T$_{\rm bb}$ decreases, consistent with what is observed.

\subsection{The distance to GX\,17+2 and the persistent emission}

Assuming that during the expansion and contraction phase the net-burst luminosity equals the
Eddington luminosity we can estimate the distance to GX\,17+2.
As noted before (Lewin et al.\ 1993; see also recent discussion by Kuulkers \&\ van der Klis 2000), 
this is not without problems. Assuming (see Appendix B) 
a neutron star with a canonical mass of 1.4\,M$_{\sun}$,
using the electron scattering opacity, $\kappa$, 
for cosmic composition in the low-temperature limit, 
we obtain L$_{\rm Edd}=4\pi cGM/\kappa=2\times 10^{38}$\,erg\,s$^{-1}$.
For the observed values of F$_{\rm bol}$$\simeq$1.2--1.3$\times$10$^{-8}$\,erg\,s$^{-1}$\,cm$^{-2}$
this leads to distances of d$\simeq$11.4--11.9\,kpc (assuming isotropic radiation and 
no spectral hardening, i.e.\ T$_{\rm eff,\infty}$=T$_{\rm bb}$). 
Close to the neutron star, however, one still has to take into account gravitational redshift
effects. We therefore estimated the distance by solving the relevant equations, 
for the explanation of which we refer the reader to Appendix~B. 
Correcting for gravitational redshift effects, taking the same `standard' parameters as above,
using values of $\kappa$ appropriate for very high temperatures, and correcting
for spectral hardening (T$_{\rm bb}$/T$_{\rm eff,\infty}$=1.7) we found  
distances between 10.8--12.0\,kpc, with an average of d$\simeq$11.3\,kpc 
(again see Appendix B for details). If the burst emission is highly
anisotropic ($\xi$$\simeq$2) we derive d$\simeq$8\,kpc, see below. We note that these estimates 
have systematic uncertainties of the order of 30\%\ or so (see Lewin et al.\ 1993; 
see also Kuulkers \&\ van der Klis 2000, and Appendix~B). 
At touch down the radius of the photosphere, R, presumably equals the radius of the neutron star surface.
We find values of R$\simeq$12--20\,km, where the range arises from the
different assumptions in deriving these radii (Appendix B). 

One can also use the radius expansion events to estimate
the persistent luminosity in Eddington units, without the need of knowing the
distance to the source, mass of the neutron star, etc. However, for the long bursts
we find that the total emission before/after the burst 
is a up to factor of $\sim$2 larger than the net-burst emission during expansion/contraction phase.
If indeed the maximum net-burst flux equals the Eddington flux, this implies
that the persistent flux is up to 2 times the Eddington flux. From the models characterising
the behaviour of Z sources it is inferred that on the HB, NB and FB, 
the mass accretion rates (and therefore luminosity) are just below, at or just above Eddington values,
respectively (e.g.\ Hasinger 1987; Lamb 1989; Hasinger et al.\ 1990). 
The only way out is that the net burst emission is (highly) anisotropic compared to
the persistent emission. Since the true luminosity values should be more or less comparable,
one thus infers an anisotropy factor of $\sim$2 for the net burst emission.
This cannot be ascribed to localized emission on the neutron star (see also Section 5.4), but
might be due to part of the burst emission being hidden behind a puffed-up inner accretion disk.
In the case of highly anisotropic burst emission ($\xi$$\simeq$2) the distance estimate to GX\,17+2 is
reduced to $\simeq$8\,kpc (see Appendix~B). This estimate is comparable to that derived for other
similar X-ray sources (e.g.\ Christian \&\ Swank 1997; Schulz 1999).
We note that anisotropic burst emission may also explain the apparent inconsistency in the distance to
Cyg\,X-2 from a radius expansion burst, i.e.\ $\simeq$11.6\,kpc (Smale 1998; 
for a 1.9\,M$_{\sun}$ neutron star, peak photospheric radius of 26\,km, cosmic 
abundances, low temperature opacity, no spectral hardening; see also Appendix B), and that 
derived from optical measurements, i.e.\ $\simeq$7.2\,kpc (Orosz \&\ Kuulkers 1999). 
However, for Cyg\,X-2 the persistent flux before the burst was of the same order as
the net peak burst flux. 

\subsection{X-ray bursts and (extreme) mass accretion rates}

We found no significant correlation between the burst parameters
and the position of the source in the Z-track at the time of the bursts, i.e.\ both long and short bursts
may occur at similar mass accretion rates. 
We find that no X-ray bursts occurred when the source was on the FB, despite
considerable coverage compared to the other branches of the Z. However, this 
result is not sufficiently significant to exclude the possibility that
bursts occur on the FB as frequently as elsewhere. We do note that the absence of bursts
on the FB is consistent with
theoretical expectations (see Sect.~1; e.g.\ Bildsten 1998, 2000), 
as the mass accretion rate on the FB is thought to be super-Eddington (e.g.\ Hasinger 1987; Lamb 1989; 
Hasinger et al.\ 1990).

At near-Eddington accretion rates, long bursts are expected from 
X-ray burst theory (see Sect.~1; see also van Paradijs et al.\ 1988). These are thought to be 
due to mixed H/He burning, triggered by thermally unstable He ignition. 
The fast rise and the short radius expansion part of the long bursts resemble
pure helium flashes which last on the order of 5--10\,s, although in the case of
GX\,17+2 the recurrence times of these bursts are on the order of a day instead
of the expected hours for typical pure helium flashes. 
This flash triggers the long phase of unstable mixed H/He burning.
Our analysis provides the first example of the existence of such long bursts at 
high (near-Eddington) accretion rates. However, GX\,17+2 also displays short 
($\simeq$10\,s) bursts, which are {\it not} expected at high accretion rates.
Similarly, Cyg\,X-2, also accreting at near-Eddington values, displays rather short
bursts ($\simeq$5\,s). Such short bursts (presumably He flashes) 
are only expected for accretion rates which are a factor of 20--100 lower.

Since it is not the global mass accretion rate that matters, but the mass accretion rate 
{\it per unit area} (e.g.\ Marshall 1982; see also Bildsten 2000), it might be that at different times
the area which accumulates most of the accreted matter differs, giving rise to short
and long bursts. However, if the areas are relatively small, pulsations are expected
due to the neutron star rotation 
(unless the accreted matter is distributed symmetrically along the rotation axis, such as
in an equatorial belt). We do not see any evidence for this.

The properties of X-ray bursts depend not only on the mass accretion rate
(per unit area), but also on the temperature of the neutron star envelope and
composition of the accreted material (e.g.\ Fushiki \&\ Lamb 1987; Taam et al.\ 1996;
see also Bildsten 1998, 2000; Lamb 2000). 
At high accretion rates and high envelope temperatures
the combined He flash and mixed H/He burning occurs; however at high accretion rates and 
low envelope temperatures only He flashes may occur (see Lamb 2000). 
The difference in envelope temperature at different times may indeed explain 
the occurrence of the two types of bursts, and it would be worthwhile to explore this further.
It is not clear, however, how such different envelope temperatures could be 
reached at very similar accretion rates. The composition of the accreted material is 
not expected to change much, since fresh matter arrives at high rates.

The situation at high mass accretion rates becomes even more
confusing when one considers the bright `GX atoll' sources 
(GX\,3+1, GX\,13+1, GX\,9+1, GX\,9+9) as well as the other Z sources 
(Sco\,X-1, GX\,5$-$1, GX\,340+0, GX\,349+2, Cyg\,X-2).
The former are thought to accrete with rates around 10\%\
of the Eddington mass accretion rate (Psaltis \&\ Lamb 1998; see also 
Kuulkers \&\ van der Klis 2000). For those sources, only 
short ($\simeq$10--15\,s duration) bursts are seen 
(infrequently) in GX\,3+1 (Makishima et al.\ 1983; Asai et al.\ 1993; 
Molkov et al.\ 1999; Kuulkers \&\ van der Klis 2000) and GX\,13+1 (Matsuba et al.\ 1995) and
no bursts have been reported for GX\,9+1 and GX\,9+9, despite ample observing times.

The X-ray spectral and fast
timing properties of the other five canonical Z sources are very similar to those of GX\,17+2
(see Hasinger \&\ van der Klis 1989; van der Klis 2000), and thus
one would infer comparable mass accretion
rates. Of the five of these sources, however,
only Cyg\,X-2 shows X-ray bursts, which are typical He flashes, whereas the others do not burst at all
(despite ample observing times). Since at very high accretion rates no bursts are expected
one would then naively infer that the average
mass accretion rate {\it is} different, i.e.\ higher, in the four non-bursting Z sources with respect
to the two bursting Z sources. But this would imply that the correlated
X-ray spectral and fast timing properties are {\it not} a function of the
inferred mass accretion rate; van der Klis (2001) has recently proposed
a way in which the correlated properties could
vary in response to changes in the mass accretion rate without being a function of it.
A slightly different suggestion was made by Homan et al.\ (2001) who
propose that the mass accretion rate is not the parameter that determines
the position along the Z track. As a consequense, bursts that occur at a
similar position along the Z might occur at different mass accretion
rates and therefore have different proporties.
Note also that differences in mass accretion rate per unit area
may be of importance here, as was already suggested by Kuulkers et al.\ (1997)
to explain the properties of quasi-periodic oscillations occurring on the HB
and upper parts of the NB, and the presence of bursts in GX\,17+2.
On the other hand, since the envelope temperatures and composition
also influence the bursting properties, they may be of importance too
(e.g.\ Taam et al.\ 1996).
Future modelling may provide more insight in the processes involved.

\section{Conclusions}

$\bullet$ We found ten X-ray bursts in all data on GX\,17+2 obtained with the RXTE/PCA to date.
Three of them were short ($\simeq$10\,s), while the others lasted
from $\simeq$6 to 25\,min. All the bursts showed spectral softening during the decay, indicative 
of cooling of the neutron star. No evidence for high-frequency
($>$100\,Hz) oscillations at any phase of the bursts is seen. 
We find no evidence for correlations of the burst properties with respect to the 
source position 
in the colour-colour diagram (presumably a function of the mass accretion rate).
The long bursts are consistent with being due to mixed H/He burning, triggered by thermally unstable He 
ignition, as expected at the inferred near-Eddington mass accretion rates
in GX\,17+2. However, the presence of short
bursts in GX\,17+2, as well as short bursts in Cyg\,X-2, another persistently bright LMXB, 
and {\it no} bursts in the other four similar LMXBs Sco\,X-1, GX\,5$-$1, GX\,340+0 and
GX\,349+2, is not accounted for in the current X-ray burst theories.
Note that this also holds for the bright `GX atoll' sources GX\,3+1, GX\,13+1, GX\,9+1 and GX\,9+9,
which are thought to accrete near one tenth of the Eddington rate. Of these four atoll sources, only
GX\,3+1 and GX\,13+1 infrequently show short ($\simeq$10--15\,s) X-ray bursts, whereas GX\,9+1 and GX\,9+9
have not been seen to burst at all.

$\bullet$ We found that two-component spectral fits to the total burst emission, as has been suggested
previously, do not give satisfactory results whenever the persistent emission before 
the burst contains a black-body component. On the other hand, the `standard' spectral
fit analysis, in which the burst emission is modeled by black-body radiation 
after subtraction of the persistent emission before the burst, does provide satisfactory results.
This means that whenever there is a black-body contribution in the persistent emission
before the burst, it is probably also present during the burst itself
(i.e.\ the burst black-body emission appears on top of the 
persistent black-body emission). This implies that the burst emission
does {\it not} arise at the same site as the black-body component often seen in the
persistent emission. We find evidence that the emission region of the persistent black-body component is larger
than that of the burst emission, which indicates that it
probably originates in an expanded boundary layer, as recent modeling indicates. 

$\bullet$ Five of the long bursts showed evidence
for an expanding photosphere during the first seconds of the burst, presumably due to the
burst luminosity reaching the Eddington limit. The contraction phase differs in duration between the bursts, 
the longest being $\simeq$3\,min. 
The net burst fluxes reached during the radius expansion/contraction phase are all more or less similar, 
which is indicative of the Eddington limit as observed at earth.

$\bullet$ The persistent flux just before the burst is inferred to be up to a factor 2 higher than the
net burst flux during the radius expansion/contraction phase. If in both cases
the emission is isotropic, this implies that the persistent emission is up to a
factor of 2 higher than the Eddington luminosity. 
We suggest that the burst emission is (highly) anisotropic.
Since both the persistent
luminosity in GX\,17+2 and the net burst luminosity during the expansion/contraction phase
are thought to have near Eddington values, we then derive an anisotropy factor of $\sim$2 for the
burst emission. 

$\bullet$ When the burst luminosity is near Eddington values, deviations from pure black-body 
radiation are evident below 10\,keV. Similar deviations
have been seen during (long) X-ray bursts in other low-mass X-ray binaries, and can not be
explained by spectral hardening.
Assuming that the black-body approximation is, nevertheless valid, that the 
net burst peak fluxes during the radius expansion/contraction phase equals the 
Eddington limit as seen on Earth, and using an anisotropy factor of 2, 
we estimate the distance to GX\,17+2, taking into account gravitational 
redshift effects and spectral hardening.
For `standard' parameters ($M_{\rm ns}$=1.4\,M$_{\sun}$, 
cosmic composition) we derive a distance $\sim$8\,kpc, with
a systematic uncertainty of up to $\sim$30\%.

\begin{acknowledgements}
We thank Lars Bildsten, Jean in 't Zand and Marten van Kerkwijk for 
providing comments on earlier drafts of this paper, 
Keith Jahoda for providing up-to-date response matrices, and the referee 
for carefully reading the manuscript.
EK thanks the MIT/RXTE X-ray burst group for making us aware of the burst-like
event f1, which resulted in our detection of the events f2--f4. 
This work was supported in part by the Netherlands Organization for Scientific Research 
(NWO). WHGL is grateful for support from NASA.
\end{acknowledgements}

\appendix

\section{X-ray bursts from GX\,17+2 observed with other instruments}

X-ray bursts from GX\,17+2 have been previously seen with the 
Monitor Proportional Counter (MPC) onboard the {\it Einstein Observatory}
(Kahn \&\ Grindlay 1984), the  second Fine Modulation Collimator (FMC-2) 
onboard {\it Hakucho} (Tawara et al.\ 1984c)
and the Medium Energy (ME) experiment onboard {\it EXOSAT} (Sztajno et al.\ 1986; 
Kuulkers et al.\ 1997).
Not all the bursts parameters as tabulated in Table~4 for the bursts seen with the
RXTE/PCA have been given by these
authors. In order to compare our bursts in more detail with the other observed bursts
we calculated these parameters from the information given in those papers.
Since the X-ray energy ranges in which the different instruments operate 
are comparable, and most of the (burst) emission is radiated in these
energy bands, such a comparison is feasible.
In the next subsections we describe how the parameters were determined for the 
different bursts.

\begin{table*}
\caption{Burst parameters of other bursts in the literature}
\begin{tabular}{lcccccl}
\hline
\multicolumn{1}{c}{burst} & \multicolumn{1}{c}{F$_{\rm pers}$$^a$} &
\multicolumn{1}{c}{F$_{\rm bb,max}$$^b$} &
\multicolumn{1}{c}{E$_{\rm b}$$^c$} &  \multicolumn{1}{c}{$\tau$ (s)} &
\multicolumn{1}{c}{$\gamma$} & \multicolumn{1}{c}{$\alpha$} \\
\hline
\multicolumn{7}{l}{Kahn \&\ Grindlay (1984)} \\
    & 1.56$\pm$0.02       & see text      & see text    &  6.7$\pm$1.4 &  3.4$\pm$0.2  & $\gtrsim$1750 \\
\multicolumn{7}{l}{Tawara et al.\ (1984c)} \\
A   & $\simeq$2.8         & $\simeq$1.5   & $\simeq$346 &  367$\pm$43  & 1.83$\pm$0.19 & $\sim$1000    \\
B   & $\simeq$2.8         & $\simeq$2.0   & $\simeq$94  &   75$\pm$8   & 1.38$\pm$0.13 & $\simeq$1300  \\
C   & $\simeq$2.9         & $\simeq$1.7   & $\simeq$114 &  109$\pm$37  & 1.77$\pm$0.20 & $\sim$1000    \\
D   & $\simeq$2.9         & $\simeq$2.1   & $\simeq$406 &  309$\pm$27  & 1.41$\pm$0.11 & $\sim$1000    \\
\multicolumn{7}{l}{Sztajno et al.\ (1986)} \\
{\sf I}   & 1.51$\pm$0.13 & 1.1$\pm$0.4   & 4.7$\pm$1.0 & 6.27$\pm$1.2 &  3.6$\pm$0.2  & $\gtrsim$1450 \\
{\sf II}  & 1.86$\pm$0.15 & 0.99$\pm$0.05 & 114$\pm$3   &  132$\pm$4   & 2.13$\pm$0.02 & $\gtrsim$650 \\
\multicolumn{7}{l}{Kuulkers et al.\ (1997)} \\
{\sf III} & -- & -- & --                                & $\simeq$27   & $\simeq$2.7   & $\gtrsim$7000 \\
{\sf IV}  & -- & -- & --                                & $\simeq$100  & $\simeq$3.2   & $\simeq$2200  \\
\hline
\multicolumn{7}{l}{\footnotesize $^a$\,Unabsorbed persistent flux in 10$^{-8}$\,erg\,s$^{-1}$\,cm$^{-2}$, see also text.} \\
\multicolumn{7}{l}{\footnotesize $^b$\,Peak net black-body flux in 10$^{-8}$\,erg\,s$^{-1}$\,cm$^{-2}$.} \\
\multicolumn{7}{l}{\footnotesize $^c$\,Burst fluence in 10$^{-8}$\,erg\,cm$^{-2}$.} \\
\end{tabular}
\end{table*}

\subsection{MPC/Einstein}

Kahn \&\ Grindlay (1984, see also Kahn et al.\ 1981) observed a burst from GX\,17+2 in 1980
with the MPC onboard Einstein, with a decay time of $\simeq$7.5\,s.
From their background subtracted count rate profiles in the 1.4--14.4\,keV energy
band (their figure 1) we determined the values for $\gamma$ and $\tau$.
The value of $\gamma$ is consistent with what is inferred using the quoted net peak and persistent 
count rate of 0.15 and 0.56\,Crab (2--6\,keV), respectively, by Kahn et al.\ (1981).
The persistent flux before the burst was given as 1.56$\pm$0.02$\times$10$^{-8}$\,erg\,cm$^{-2}$\,s$^{-1}$
(1--20\,keV). 
Kahn \&\ Grindlay (1984) report a maximum peak flux of 9.1$\times$10$^{-8}$\,erg\,cm$^{-2}$\,s$^{-1}$
(1--20\,keV). No error is given, but only a $\pm$-sign. It is not clear whether this is a typo, or
if this means that the value was approximate.
Their peak flux is much higher than observed for the other bursts. 
An averaged net spectrum over the burst profile (first 7.68\,s) was modeled
as black-body emission. Values of $3.8\lesssim {\rm T}_{\rm bb}\lesssim 9.3$\,keV and
black-body emission area, A$_{\rm bb}$, of 
$0.56\lesssim {\rm A}_{\rm bb}\lesssim 3.2$$\times$10$^{12}$\,cm$^2$ (at 10\,kpc) were found.
This, in principle, may give us a handle on the observed fluence. However,
this gives us a value of E$_{\rm b}$ of $\simeq$161$\times$10$^{-8}$\,erg\,cm$^{-2}$ 
(using T$_{\rm bb}$$\simeq$6.55\,keV and A$_{\rm bb}$$\simeq$1.88$\times$10$^{12}$\,cm$^2$),
which is also much larger than observed for the other short bursts. We, therefore, do not quote 
the peak flux and burst fluence in Table~A.1. 
Using the Einstein database maintained at HEASARC\footnote{See http://heasarc.gsfc.nasa.gov/.}
we found the start of the (simultaneous HRI) observation as being March 29, 1980, 00:48:06 (UTC). 
The burst was observed $\simeq$57\,min later (Kahn et al.\ 1981). Using this and the observed rate just before
the burst we determined an upper limit on $\alpha$.

\subsection{FMC-2/Hakucho}

Four bursts (two in 1981 and two in 1982; denoted A--D) were found, originating 
from GX\,17+2, during pointed observations
with the FMC-2 onboard Hakucho (Tawara et al.\ 1984c).
They had e-folding decay times of $\simeq$100\,s (B, C) or $\simeq$300\,s (A, D).
For each burst Tawara et al.\ (1984c) tabulated the net peak, net total (=fluence) and 
persistent flux in FMC-2 cts\,s$^{-1}$ in the 1--22\,keV band. From this they
calculated $\gamma$. Using this information we calculated $\tau$ and verified
$\gamma$. 
Tawara et al.\ (1984c)  also provided 
the total integrated energy fluxes (= total fluence) of the two bursts in 1981 and
the two bursts in 1982, together with the integrated persistent fluxes over 
40\,hr and 50\,hr, respectively. Note that it is not clear if the fluence and 
persistent flux are both given for the 1--22\,keV energy band, and 
whether the persistent fluxes are corrected for interstellar absorption or not. 
From this it was estimated that $\alpha$ was of the 
order of 1000. Since the net fluence in counts of all bursts individually are given, we can
determine the individual fluence in energy flux. Also, the (mean) persistent flux 
during the two observations can be determined. Moreover, assuming that between bursts
A and B, which occurred 12.5\,hr after each other, no other bursts occur, we give
a more exact value for $\alpha$ in this case (assuming a constant persistent flux). 
For the others we just assume the given value of
1000. We also give an estimate of the net peak burst flux, from the observed values
of F$_{\rm pers}$ and $\gamma$.

\subsection{ME/EXOSAT}

Two bursts were found in the EXOSAT/ME observations taken in 1984 and 1985 
with durations of $\simeq$10\,s and $>$5\,min
(Sztajno et al.\ 1986; denoted as burst {\sf I} and {\sf II} by Kuulkers et al.\ 1997).
The persistent flux (2--20\,keV) before a burst was determined from 
the spectral fits done by Sztajno et al.\ (1986), who used 
a bremsstrahlung plus a black-body component, subjected to interstellar absorption. 
The values for the bremsstrahlung parameters and N$_{\rm H}$ were given in their
text, whereas the black-body results were indicated in their figures 4 and 5.
The spectral fit results to the net burst emission are shown in their figures 2 and 3, from
which we determined the bolometric maximum net peak flux and the fluence. 
To determine the `rest' fluence, we used the same approach as mentioned in 
Sect.~3.2.5, and fitted an exponential to the burst flux decay.
In the case of burst~{\sf I} we have only 3 values; we therefore assumed that at infinity
the burst flux decays to zero, while the exponential start time was set to the 
start of the burst.
Since the spectral time resolution is not sufficient, we determined the burst parameters
$\gamma$ and $\tau$ from the observed raw (i.e.\ not dead-time corrected) 
count rate profiles (1--20\,keV) given in their figure 1. We corrected the light curves for 
the `variable' dead time (van der Klis 1989), as appropriate for the EXOSAT 
High Energy Resolution (HER) modes (see e.g.\ Kuulkers 1995).
Sztajno et al.\ (1986) noted that in both bursts the maximum flux is $\simeq$40\%\
above the persistent level, indicating $\gamma$$\simeq$2.5, in rough agreement
with our estimates. 
Using the start time of the two observations during which the bursts occurred 
as given by Kuulkers et al.\ (1997), and assuming that the persistent flux was constant, 
we were able to determine lower limits on $\alpha$. Note that Sztajno et al.\ (1986) 
gave rough upper limits of $\sim$10000 and $\sim$500, for burst {\sf I} and {\sf II}, 
respectively, which are more or less consistent with our determined values.

For the two bursts (denoted as burst {\sf III} and {\sf IV}) observed in 1986 discussed by 
Kuulkers et al.\ (1997) we only estimated
$\gamma$ and $\tau$ from the observed (dead-time and background corrected) count rate profiles 
in the 1--20\,keV band as displayed in their figure~7.
$\tau$ is determined using the integrated net burst count rate profile and peak count rate.
No spectral information is available for the persistent or burst emission in order 
to determine F$_{\rm bb,max}$, E$_{\rm b}$ and $\alpha$, although they note that
$\alpha$ is probably larger than $\sim$1000 for the second burst, if no bursts occurred
in between the two bursts. However, if we naively assume that the persistent count rate
is constant between bursts, and that between the start of the 1986 EXOSAT observation
and the first burst, and between the two bursts, no other burst occurred, we can still
get a crude estimate of (the lower limit on) the value of $\alpha$.

\section{Distance estimate and gravitational redshift effect}

\begin{table*}
\caption{Distance (d in kpc) and radius at touch down from radius expansion bursts$^a$}
\begin{tabular}{lccccccc}
\hline
\multicolumn{1}{c}{~} & \multicolumn{3}{c}{0.25\,s spectra} & 
\multicolumn{3}{c}{16\,s spectra} & \multicolumn{1}{c}{R$_{\rm td}$$^b$} \\
burst & \#$^c$ & $d_{\rm non\_rel}$$^d$ & $d_{\rm rel}$$^e$ & \# & $d_{\rm non\_rel}$ & $d_{\rm rel}$ & (km) \\
\hline
b4 & 84 & 11.4$\pm$0.3 & 10.8$\pm$0.3 & 4 & 11.6$\pm$0.3 & 10.9$\pm$0.3 & 18.3 \\
b6 & 83 & 12.1$\pm$0.5 & 11.4$\pm$0.5 & 6 & 12.2$\pm$0.1 & 11.1$\pm$0.1 & 12.8 \\
b7 & 94 & 12.9$\pm$0.6 & 12.0$\pm$0.5 & 2 & 12.7$\pm$0.2 & 11.8$\pm$0.1 & 16.7 \\
b8 & 93 & 12.3$\pm$0.6 & 11.5$\pm$0.5 & 2 & 12.1$\pm$0.1 & 11.2$\pm$0.1 & 14.4 \\
b9 & 82 & 12.4$\pm$0.5 & 11.5$\pm$0.5 & 1 & 12.1         & 11.1         & 14.9 \\
\multicolumn{2}{l}{Average:} & 12.2$\pm$0.2 & 11.4$\pm$0.2 &   & 12.1$\pm$0.1 & 11.2$\pm$0.1 & 15.4$\pm$1.0 \\
\hline
\multicolumn{8}{l}{\footnotesize $^a$\,For the parameters used we refer to the text.}\\
\multicolumn{8}{l}{\footnotesize $^b$\,Black-body radius R at touch down for a local observer (see text).}\\
\multicolumn{8}{l}{\footnotesize $^c$\,Number of spectra in the expansion/contraction phase.}\\
\multicolumn{8}{l}{\footnotesize $^d$\,Not taking into account gravitational redshift effects.}\\
\multicolumn{8}{l}{\footnotesize $^e$\,Taking into account gravitational redshift effects.}\\
\end{tabular}
\end{table*}

Most of the distances derived from radius expansion bursts in the literature are solely based
on the observed fluxes at maximum expansion, F$_{\rm bb,max}$, and assuming this equals the
(non-relativistic) Eddington luminosity, L$_{\rm Edd,non\_rel}$, from a source at distance $d$. Here one
uses L$_{\rm Edd,non\_rel}=4\pi cGM_{\rm ns}/\kappa$, where $c$ is the speed of light,
$G$ the gravitational constant, $M_{\rm ns}$ the mass of the neutron star and 
$\kappa$ the electron scattering opacity. Then the distance may be estimated as:
\begin{equation}
d = \left( \frac{{\rm L}_{\rm Edd,non\_rel}}{4\pi \xi {\rm F}_{\rm bb,max}} \right)^{1/2},
\end{equation} 
where $\xi$ is the anisotropy factor (for isotropic radiation $\xi$=1).
However, when the photosphere remains rather close to the
neutron star, gravitational redshift effects become important (due to relativistic time dilation,
see e.g.\ Lewin et al.\ 1993).
The distance derived using Eq.~B.1 is in fact only valid for values of 
the photosphere radius, R$\gg$R$_{\rm ns}$ (see below). For R$\gtrsim$R$_{\rm ns}$
Eq.~B.1 in principle only provides an upper limit on $d$.

The (relativistic) Eddington luminosity for an observer near the neutron star 
is given by:
\begin{equation}
{\rm L}_{\rm Edd}=(4\pi cGM_{\rm ns}/\kappa)[1-2GM_{\rm ns}/({\rm R}c^2)]^{-1/2}.
\end{equation}
For an observer at Earth, Eq.~B.2 becomes, taking into gravitational redshift effects:
\begin{equation}
{\rm L}_{\rm Edd,\infty}={\rm L}_{\rm Edd}[1-2GM_{\rm ns}/({\rm R}c^2)] = 
{\rm L}_{\rm Edd}[1-{\rm R}_g/{\rm R}],
\end{equation}
where R$_g$ is the Schwarzschild radius (${\rm R}$$\geq$1.5R$_g$; see Lewin et al.\ 1993).
Also the effective black-body temperature measured at Earth, T$_{\rm eff,\infty}$, 
is affected by the gravitational potential:
\begin{equation}
{\rm T}_{\rm eff,\infty} = {\rm T}_{\rm eff}[1-{\rm R}_g/{\rm R}]^{1/2},
\end{equation}
where T$_{\rm eff}$ is the effective black-body temperature 
near the neutron star.
Assuming that the source radiates as a black-body, we know that:
\begin{equation}
{\rm L}_{\rm Edd}=4\pi {\rm R}^2\sigma T_{\rm eff}^4.
\end{equation}
For L$_{\rm Edd,\infty}$ we also have:
\begin{equation}
{\rm L}_{\rm Edd,\infty}=4\pi d^2\xi {\rm F}_{\rm Edd,\infty}, 
\end{equation}
where F$_{\rm Edd,\infty}$ is the Eddington flux received at Earth.

Eqs.~B.3 and B.4 are strictly valid only for non-rotating stars, but
it is still approximately correct for stars with rotation periods of a few milliseconds.
F$_{\rm Edd,\infty}$ is set equal to the observed net-burst fluxes, F$_{\rm bb,net}$, 
during the radius expansion/contraction phase. 
As noticed in Sect.~5.2, it is not easy to infer T$_{\rm eff,\infty}$ from our observed values of
T$_{\rm bb}$. We therefore fixed the hardening factor, T$_{\rm bb}$/T$_{\rm eff,\infty}$, to different values
in order to assess its effect. We also assume certain values for $M_{\rm ns}$ and $\xi$, and
are then left with five equations (B.2--B.6) and five unknown variables 
(L$_{\rm Edd}$, L$_{\rm Edd,\infty}$, T$_{\rm eff}$, R and d); this enables us to derive a distance
estimate.
First, one can eliminate L$_{\rm Edd}$ and T$_{\rm eff}$ by combining
Eqs.~B.2, B.4 and B.5, and one gets the following equality (for R$\geq$1.5R$_g$):
\begin{equation}
\frac{cGM_{\rm ns}}{\kappa\sigma {\rm T}_{\rm eff,\infty}^4} = {\rm R}^2\left( \frac{\rm R}{{\rm R}-{\rm R}_g} \right)^{\frac{3}{2}}.
\end{equation}
Since this equation can not be solved easily in an analytic way, we solved for R numerically. 
Using R we can then determine L$_{\rm Edd,\infty}$ (Eq.~B.3), 
which in turn can be used to determine the distance, $d$ (Eq.~B.6).

In the low-temperature limit
the electron scattering opacity, $\kappa$, is given by $\kappa=0.2(1+X)$\,cm$^2$\,g$^{-1}$, where $X$ is the 
hydrogen fraction (by mass) of the photospheric matter (see e.g.\ Lewin et al.\ 1993). 
Note that for cosmic compositions $X$=0.73. 
At very high temperatures (i.e.\ probably near the peak of bursts), 
however, the scattering electrons become relativistic
and $\kappa$ may instead be approximated by (in the low-density
limit) $\kappa=\kappa_0[1+({\rm kT}/39.2~{\rm keV})^{0.86}]^{-1}$, where 
$\kappa_0$ is same as for the low-temperature limit (Paczy\'nski 1983).
It is not entirely clear how kT relates to our observed values of kT$_{\rm bb}$,
but one may assume kT=kT$_{\rm bb}$ (e.g.\ van Paradijs et al.\ 1990).
We use the high temperature electron scattering opacity in our calculations.
The estimated distances differ by a factor of $\simeq$1.05 with respect to
the estimates when taking the electron scattering opacity in the low-temperature limit
(see also Table~B.2).

The above described procedure for estimating the distance was followed using the 
values for F$_{\rm bb,net}$ and T$_{\rm bb}$ for each of the spectral fits to
the 0.25\,s and 16\,s spectra in the radius expansion/contraction phase.
When we assume no spectral hardening, i.e.\ T$_{\rm eff,\infty}$=T$_{\rm bb}$,
and `standard' parameters ($M_{\rm ns}$=1.4\,M$_{\sun}$, $X$=0.73)
no solutions can be found for those spectra having kT$_{\rm bb}$$\gtrsim$1.75\,keV,
i.e.\ only at the largest expansion radii do we have 
solutions (see also Table~B.2)\footnote{Note 
that the gravitational redshift corrected distance estimates by 
Smale (1998: Cyg\,X-2), Kuulkers \&\ van Klis (2000: GX\,3+1) and Kaptein et al.\ (2000: 1RXS\,J171824.2$-$402934)
are derived using T$_{\rm eff,\infty}$=T$_{\rm bb}$ and R=R$_{\rm bb}$,
thus only using
Eqs.~B.2, B.3 and B.6.}.
The highest observed values of kT$_{\rm bb}$ during the radius expansion/contraction phase 
give the strongest constraints on the spectral hardening factor. This is at the moment of
`touch-down', i.e.\ at the vertex of the horizontal track and diagonal track in the
flux versus temperature diagram. This locus is well defined for our long bursts observed in
GX\,17+2. Using the loci of the radius expansion bursts b4, b6--b9, we find 
that the minimum value of the (constant) hardening factor is between
1.34 (burst b4) and 1.52 (burst b6).
The distance and radius R as inferred at the loci both increase for 
increasing spectral hardening (e.g.\ for burst b6, the distance increases by a factor $\simeq$1.1, 
while R increases by a factor of $\simeq$2 when the spectral hardening factor increases
from 1.6 to 2). 

In Table~B.1 the results for `standard' burst parameters can be seen, i.e.\ $M_{\rm ns}=1.4$\,M$_{\sun}$,
cosmic composition of the photospheric matter, isotropic radiation ($\xi$=1), as well
as taking T$_{\rm bb}$/T$_{\rm eff,\infty}$=1.7 and using the electron scattering opacity appropriate
for high temperatures.
We show the distance as derived by including gravitational redshift effects ($d_{\rm rel}$) and
not including these effects ($d_{\rm non\_rel}$, i.e.\ using Eq.~B.1). 
The errors are set equal to the observed variances in the derived distances.

\begin{table*}
\caption{Distance ($d$ in kpc) and radius at touch down for burst b6 using different assumptions$^a$}
\begin{tabular}{cccccc}
\hline
\multicolumn{1}{c}{~} & \multicolumn{2}{c}{0.25\,s spectra} & 
\multicolumn{2}{c}{16\,s spectra} & \multicolumn{1}{c}{R$_{\rm td}$$^b$} \\
parameter$^a$ & $d_{\rm non\_rel}$$^c$ & $d_{\rm rel}$$^d$ & $d_{\rm non\_rel}$ & $d_{\rm rel}$ & (km) \\
\hline
standard$^e$                   & 12.1$\pm$0.5 & 11.4$\pm$0.5 & 12.2$\pm$0.1 & 11.1$\pm$0.1 & 12.8 \\
$M_{\rm ns}$=2.0\,M$_{\sun}$  & 14.4$\pm$0.6 & 13.5$\pm$0.6 & 14.6$\pm$0.1 & 12.7$\pm$0.2 & 13.0 \\
$X$=0                          & 15.9$\pm$0.7 & 15.3$\pm$0.6 & 16.1$\pm$0.1 & 15.1$\pm$0.1 & 18.9 \\
low-T $\kappa$                 & 11.6$\pm$0.5 & 11.0$\pm$0.5 & 11.7$\pm$0.1 & 10.6$\pm$0.1 & 11.8 \\
$\xi$=2                        &  8.5$\pm$0.3 &  8.1$\pm$0.3 &  8.6$\pm$0.1 &  7.9$\pm$0.1 & 12.8 \\
$\xi$=0.5                      & 17.1$\pm$0.7 & 16.2$\pm$0.7 & 17.3$\pm$0.1 & 15.8$\pm$0.1 & 12.8 \\
T$_{\rm bb}$/T$_{\rm eff,\infty}$=2   & 12.1$\pm$0.5 & 11.7$\pm$0.5 & 12.2$\pm$0.1 & 11.6$\pm$0.1 & 19.9 \\
T$_{\rm bb}$/T$_{\rm eff,\infty}$=1$^f$ & 12.1$\pm$0.5 (83) & 10.5$\pm$0.7 (9) & 12.2$\pm$0.1 (6) & --- & --- \\
\hline
\multicolumn{6}{l}{\footnotesize $^a$\,See text for details.}\\
\multicolumn{6}{l}{\footnotesize $^b$\,Black-body radius R at touch down for a local observer (see text).}\\
\multicolumn{6}{l}{\footnotesize $^c$\,Not taking into account gravitational redshift effects.}\\
\multicolumn{6}{l}{\footnotesize $^d$\,Taking into account gravitational redshift effects.}\\
\multicolumn{6}{l}{\footnotesize $^e$\,$M_{\rm ns}$=1.4\,M$_{\sun}$, $X$=0.73, T$_{\rm bb}$/T$_{\rm eff,\infty}$=1.7,high-T $\kappa$, $\xi$=1.}\\
\multicolumn{6}{l}{\footnotesize $^f$\,Between brackets the number of spectra taken into account are given.}\\
\end{tabular}
\end{table*}

By relaxing the assumptions of standard burst parameters (i.e.\ varying $M_{\rm ns}$, $X$
and $\xi$ individually within reasonable limits, while keeping the others at the 
`standard' values) one can get an estimate of the systematic uncertainties
involved (see e.g.\ also Kuulkers \&\ van der Klis 2000).
The results are shown in Table~B.2 for burst b6, which is the longest burst of our sample. 
We show the results assuming a neutron star mass of $M_{\rm ns}=2$\,M$_{\sun}$, in the range required 
in models for explaining the kHz QPO in neutron star LMXBs
(see e.g.\ Stella et al.\ 1999; Stella 2001; Lamb \&\ Miller 2001),
and close to the dynamical neutron mass estimate of Cyg\,X-2 using optical
spectroscopy/photometry, $M_{\rm ns}=1.78$$\pm$0.23 (Orosz \&\ Kuulkers 1999).
Note that the distance derived is rather high (Table~B.2).
We also made the assumption 
that the photospheric composition is hydrogen-poor (i.e.\ $X$=0), as has been argued to
be the case during radius expansion (see Sugimoto et al.\ 1984).
However, the contraction phase of long bursts are typical of unstable mixed hydrogen/helium
burning (see Sect.~5.4), which suggests that this is not a reasonable assumption.
Also, the persistent mass accretion rate is inferred to be high, making it 
plausible that the photosphere will be continuously supplied with new hydrogen matter.
Moreover, the distances derived are rather high (see Table~B.2).
We changed the anisotropy factor between the reasonable values $0.5\lesssim \xi \lesssim 2$
(see van Paradijs \&\ Lewin 1987; Lewin et al.\ 1993).
Finally, we compared the distance values derived using values for $\kappa$ valid for 
low temperatures.
The results shown in Table B.2 suggests that the systematic uncertainties on the derived distance are 
up to $\sim$30\%.

\end{document}